\documentclass{CVM}

\newcommand{\ie}{\emph{i.e.}}

\newcommand{\eg}{\emph{e.g.}}

\newcommand{\holens}{\emph{HoLens}}
\newcommand{\sect}{Sect.}
\newcommand{\fig}{Fig.~}

\newcommand{\alg}{Algorithm.~}

\newcommand{\fzz}[1]{{\color{black}{#1}}}
\newcommand{\whj}[1]{{\color{black}{#1}}}

\CVMsetup{
type      = {Research/Review Article},
doi       = {s41095-0xx-xxxx-x},
title     = {\emph{HoLens}: A Visual Analytics Design for Higher-order Movement Modeling and Visualization},
author    = {Zezheng~Feng$^{1,2}$,~Fang Zhu$^{2}$,~Hongjun~Wang$^{2}$,~Jianing~Hao$^{3}$,~Shuang-Hua~Yang$^{5,4,2}$\cor{},~Wei~Zeng$^{3}$,~Huamin~Qu$^{1}$\\
},
runauthor = {Z. Feng, F. Zhu, H. Wang, J. Hao, SH. Yang, W. Zeng, H. Qu},
abstract  = {
Higher-order patterns reveal sequential multistep state transitions, which are usually superior to origin-destination analysis, which depicts only first-order geospatial movement patterns.
Conventional methods for higher-order movement modeling first construct a directed acyclic graph (DAG) of movements, then extract higher-order patterns from the DAG.
However, DAG-based methods heavily rely on the identification of movement keypoints that are challenging for sparse movements and fail to consider the temporal variants that are critical for movements in urban environments.
To overcome the limitations, we propose \holens, a novel approach for modeling and visualizing higher-order movement patterns in the context of an urban environment.
\holens~mainly makes twofold contributions:
first, we design an auto-adaptive movement aggregation algorithm that self-organizes movements hierarchically by considering spatial proximity, contextual information, and temporal variability; second, we develop an interactive visual analytics interface consisting of well-established visualization techniques, including the H-Flow for visualizing the higher-order patterns on the map and the higher-order state sequence chart for representing the higher-order state transitions. Two real-world case studies manifest that the method can adaptively aggregate the data and exhibit the process of how to explore the higher-order patterns by \holens.
We also demonstrate our approach's feasibility, usability, and effectiveness through an expert interview with three domain experts.

},
keywords  = {Data Visualization, Movement Modeling, State Sequence Visualization, Movement Visualization, Urban Visual Analytics.

},
copyright = {The Author(s)},
}

\begin{document}

\maketitle

    \begin{figure}[b] \vskip -4mm
    \small\renewcommand\arraystretch{1.3}
        \begin{tabular}{p{80.5mm}} \toprule\\ \end{tabular}
        \vskip -4.5mm \noindent \setlength{\tabcolsep}{1pt}
        \begin{tabular}{p{3.5mm}p{80mm}}
    $1\quad $ & Hong Kong University of Science and Technology. Email: zfengak@connect.ust.hk, huamin@cse.ust.hk\\
    $2\quad $ & Southern University of Science and Technology. E-mail: \{11711623, wanghj2020\}@mail.sustech.edu.cn\\
    $3\quad $ & Hong Kong University of Science and Technology (Guangzhou). E-mail: jhao768@connect.hkust-gz.edu.cn, weizeng@ust.hk\\
    $4\quad $ & Department of Computer Science, University of Reading.\\
    $5\quad $ & Shenzhen Key Laboratory of Safety and Security for Next Generation of Industrial Internet, Southern University of Science and Technology\\
    \cor{} & ~Shuang-Hua Yang is the corresponding author. E-mail: yangsh@sustech.edu.cn; Shuang-hua.yang@reading.ac.uk\\
  
&\hspace{-5mm} Manuscript received: 2022-01-01; accepted: 2022-01-01\vspace{-2mm}
    \end{tabular} \vspace {-3mm}
    \end{figure}

\section{Introduction}\label{sec: holens-intro}

\begin{figure}[htbp]
\centering
  \includegraphics[width=\linewidth]{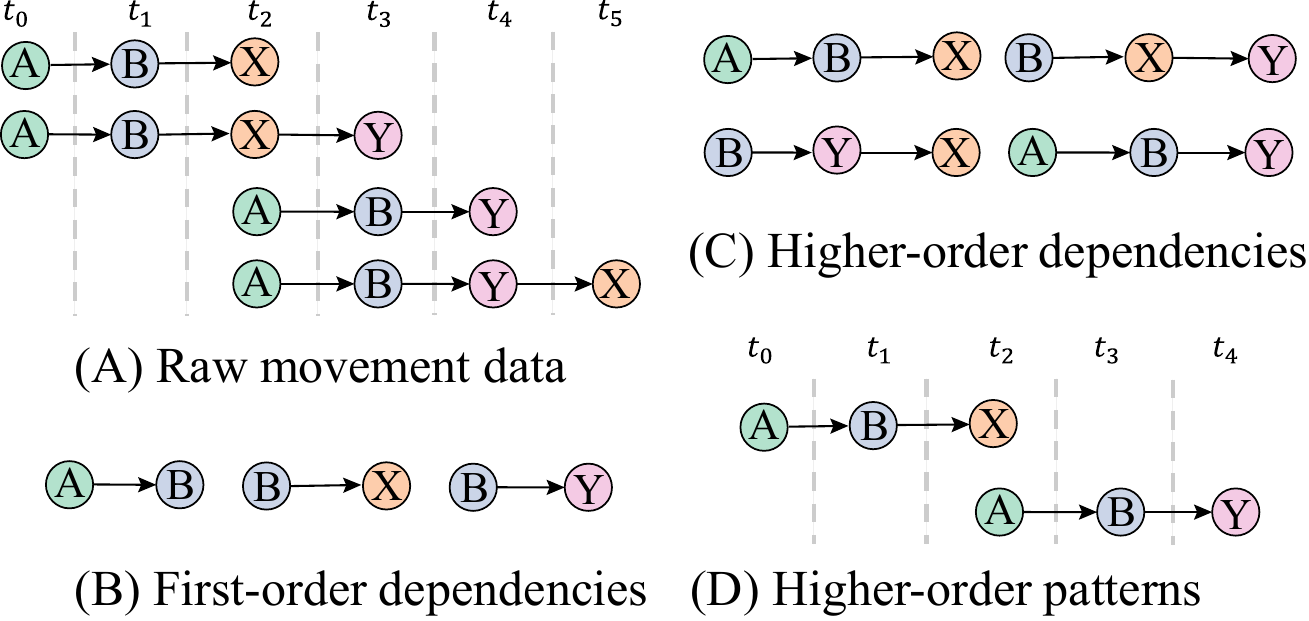}
    \vspace{-2em}
    \caption{Illustration of higher-order patterns: (A) Raw movement data describe the change of positions over time, modeled as trajectories. We can extract (B) first-order and (C) higher-order dependencies from these movement data. Moreover, a set of first-/higher-order dependencies can form (D) higher-order patterns.}
    \label{fig: holens-intro higher-order}
  \vspace{-0.5em}
\end{figure}

Many complex systems use network structures to represent interactions between entities, \eg, to construct a global shipping network to represent ship movements between ports~\cite{tao2017honvis}, or to build a neuron graph to simulate neuron transmission in the brain~\cite{xu2016representing}.
Each entity is represented as a node of the network, and flows are encoded as edges.
Such a network model implicitly assumes that the current status is dependent on only its precedent, \ie, first-order dependency as in a Markov process~\cite{markov1954theory}, ~\fig\ref{fig: holens-intro higher-order}($B$).
However, the model cannot represent scenarios where the current status may not entirely depend on its first-order precedent.
For example, when humans browse the Internet, subsequent mouse-clicking behavior may not depend on only one previous behavior~\cite{chierichetti2012web}.

Higher-order dependency analysis, which can be traced back to Shannon's high-order memory model~\cite{shannon1948mathematical}, can alleviate the problem.
The analysis is crucial to many real-world applications \eg, the analysis of animal behavior~\cite{grundy2009visualisation}, rumor spread~\cite{nekovee2007theory}.
For example, biologists gathered the data from sensors mounted on animals and extracted higher-order dependencies to analyze behavior patterns and activities~\cite{kareiva1983analyzing,grundy2009visualisation}.
By tracing multiple higher-order pathways in global ship data, species invasions are investigated and predicted~\cite{tao2017honvis}.


\fzz{When analyzing geospatial movements, higher-order patterns that reveal sequential multi-step state transitions may depict insights different from origin-destination (OD) patterns. For example, three OD patterns (\fig\ref{fig: holens-intro higher-order}($B$)) are retrieved from the raw movements in \fig\ref{fig: holens-intro higher-order}($A$) can show the relations between two states, whereas higher-order patterns (\fig\ref{fig: holens-intro higher-order}($C$)) can identify more detailed sequential relations. Additionally, in urban areas, analyzing the moving objects' multistep behavior is helpful~\cite{andrienko2018state}. For example, urban planners can plan public facilities and manage traffic when they understand the multistep moving behavior of the citizens; epidemiologists can trace back the possible propagation path and forecast the potential routes when making sense of the citizens' multistep moving pattern~\cite{andrienko2012visual}. These works heavily rely on identifying the functional attributes of each state (\ie, what do citizens do in this area) before forming higher-order patterns. Although clustering-based (\eg, \cite{gaffney1999trajectory, gaffney2007probabilistic, lee2007trajectory}) and aggregation-based techniques (\eg,~\cite{andrienko2008spatio,adrienko2010spatial, zhou_2019_visual}) have been proposed for keypoint identification, the methods may not be in line with contextual information that depicts underlying mechanisms for the movements. Furthermore, existing visual analytics have been proposed to explore higher-order movement patterns, \eg,~\cite{blaas_2009_smooth, rosvall2014memory}.
However, these studies seldom consider the influence of temporal attributes on higher-order movement patterns.
As an example, \fig\ref{fig: holens-intro higher-order}($D$) illustrates that two higher-order patterns exist at different periods, \ie, $A$$\rightarrow$$B$$\rightarrow$$X$ in period [\emph{$t_0-t_2$}] \emph{vs.} $A$$\rightarrow$$B$$\rightarrow$$Y$ in period [\emph{$t_2-t_4$}].}


This work focuses on exploring higher-order movement patterns, specifically in the context of an urban environment.
We distill a list of design considerations as follows:
\begin{itemize}[leftmargin=*]

\fzz{\item \textbf{Contextual Information.}
~Contextual information is vital for reasoning the underlying mechanisms when analyzing movements because they are related to many factors, \eg, activity distributions and household demographics.

\item \textbf{Multi-scale Self-organization.}~The movements shall be aggregated adaptively according to the static contextual information and dynamic movements. Moreover, the aggregation method shall support consistency when exploring under different scales.

\item \textbf{Temporal Variability.}~Higher-order movement patterns in an urban area are highly regular and temporal variability~\cite{zeng_2017_visualizing}, \eg, people commute to business districts in the morning and back to residential districts in the evening.}

\end{itemize}
We introduce a visual analytics approach, namely, \textbf{H}igher-\textbf{o}rder \textbf{Lens} (\holens).
\holens~makes twofold contributions in terms of modeling and visualizing higher-order movement patterns in urban areas.
First, we design a self-organization aggregation method that forms region clusters by considering the spatial proximity of the movements, contextual information, and access frequency to each location.
We then construct a directed acyclic graph (DAG) with centroids of cluster regions as nodes and interactions between per-pair cluster regions as edges.
In this manner, we can extract higher-order movement patterns, considering the spatial and temporal characteristics.
Second, we propose a multiview visual analytics system that enables the exploration of higher-order movement patterns from the spatial and temporal dimensions.
The effectiveness of \holens~is demonstrated through case studies conducted on real-world movements in New York City and through expert feedback.

The main contributions of this work are as follows:
\begin{itemize}
\item a spatial proximity- and contextural-based self-organizing method for adaptively aggregating urban movement data at different scales, which can avoid the spareness of the urban movement data;

\item an interactive visual analytics approach with a newly developed user interface named \holens,~consisting of a set of newly designed visualizations to assist analysts in exploring the higher-order patterns in urban areas;

\item two real-world case studies and expert interviews to demonstrate the feasibility, usability, and effectiveness of our proposed approach. 
\end{itemize}


\section{Related Work}\label{sec: holens-related}
Below, we summarize related studies in movement modeling and visualization (\sect\ref{sec: holens-related-movement modeling}), state sequence~(\sect\ref{sec: holens-related-state sequence}), and higher-order dependency visualization~(\sect\ref{sec: holens-related-higher-order}).

\subsection{Movement Modeling and Visualization}\label{sec: holens-related-movement modeling}
Movement data describe the change of positions over time for moving objects~\cite{dodge_2008_towards}.
Understanding movement pattern is important in many domains, \eg, animal ecology~\cite{slingsby_2016_exploratory}, social media~\cite{chen2015interactive}, and urban transportation~\cite{chen_survey_2015}.
Andrienko et al.~\cite{andrienko_2008_geovisualization} categorized three approaches for exploratory and analytical visualization of movement data: \emph{direct depiction}, \emph{summaries}, and \emph{pattern extraction}.
\emph{Direct depiction} of each movement record (\eg,~\cite{kapler_2004_geotime}) can facilitate the extraction of noteworthy patterns.
However, the approach can easily cause cluttering issues.
Appropriate \emph{summaries} can address the issues using spatial generalization and aggregation methods (\eg,~\cite{andrienko_spatial_2011-1, guo_2014_origin-destination}) or visualization methods (\eg,~\cite{scheepens_2011_composite, feng_2021_topology}).

Nevertheless, due to the dynamic and heterogeneous properties, \emph{direct depictions} or \emph{summaries} of a large amount of movement data are nontrivial.
A more general pipeline is to first conduct \emph{pattern extraction} on movement data, followed by visualization of the movement patterns.
Advanced data mining techniques (\eg,~\cite{giannotti2007trajectory, mining_pattern}) can be applied for pattern extraction.
Lee et al.~\cite{lee2007trajectory} proposed a partition-and-group framework for clustering trajectories based on ordinary trajectory clustering algorithms to find common sub-trajectories.
These methods focus on localized movement patterns, whilst neglecting the higher-order connections among the regions.
~Zeng et al.~\cite{zeng2017visual} showed that different movement rhythms are derived when the order of movement pattern is increased.
Alternatively, movements can be modeled as graphs~\cite{deng2023survey} and graph-based methods can be employed to extract higher-order patterns (\eg,~\cite{huang_2016_trajgraph, zhou_2019_visual, zeng_2019_raeb}), but these works usually neglect considering contextual information and fail to support adaptive multi-scale modeling.


In this study, we propose a novel dynamic and adaptive movement modeling method that considers both spatial proximity and contextual information of urban movement data.

\subsection{State Sequence Visualization}\label{sec: holens-related-state sequence}
State sequence visualization aims to analyze behavior that generally exhibits some form of symmetry and regularity, \eg, complex computer-based systems~\cite{visofstatetran} and chess playing~\cite{lu_2014_chess}.
The basic approach to state sequence visualization is to place events along a horizontal time axis as done by Lifelines~\cite{plaisant1996lifelines} and CloudLines~\cite{cloudlines}. However, the huge amount of nodes and edges hinders analysis and insight discovery.
To reduce the load, a common solution is to aggregate state sequences first, and then use visualizations like Sankey diagram~\cite{schmidt_2008_sankey}.
Other studies (\eg,~\cite{ham_2002_interactive, wongsuphasawat2011lifeflow, eventflow, shen2012visual, coreflow}) consolidate common subsequences, extract and visualize tree-like representation from state sequence data.
These representations may ignore some low-frequency states in the visual summary. 




Visualizing movement data as a state sequence is more challenging than conventional state sequence visualization since it needs to consider the position as a constraint.
Simply using graphs to visualize state transition sequences may cause the spatial context invisible.
Nevertheless, researchers have come up with new visual designs to depict state sequences with spatial information, \eg, interchange circos diagram for urban traffic~\cite{zeng_2013_visualizing}, state transition graphs in observational time-series\cite{blaas_2009_smooth}, and designs for animal movements~\cite{grundy2009visualisation,slingsby2016exploratory,ware2006visualizing}.
However, the methods cannot show a higher-order state and the increasing nodes will lead to visual complexity.

A novel visualization design named \emph{higher-order state sequence chart} is developed to avoid visual clutter meanwhile helping analyze higher-order patterns.
The design can express both characteristics of higher-order state sequences and temporal features of higher-order movement pattern.

\subsection{Higher-order Dependency Visualization}\label{sec: holens-related-higher-order}

Higher-order dependency, where a sequence of the preceding status exerts influence on the present state, is prevalent in complex systems, \eg, object movements and web clickstreams~\cite{xu2016representing}. In the higher-order dependency, the historical sequence is considered~\cite{zhang2016efficient}. 

Some conventional solutions consolidate the states into a graph. The state transition diagram~\cite{ham_2002_interactive} shows all routes between nodes. Lu et al.~\cite{areachart} visualize the sentiment trend of time series with stacked area charts. However, these approaches can not scale well to highly-connected graphs. Sankey diagrams~\cite{riehmann2005interactive} is the choice for many studies~\cite{outflow, decisionflow, perer2013data} to show the transition pathways. 
In addition, matrix-based representations~\cite{zhao2015matrixwave, perer2012matrixflow} are designed to avoid visual clutter caused by dense edges in Sankey diagrams. However, matrix-based representation introduces usability issues.

For interactive exploration, Chen et al.~\cite{chen2015interactive} developed a visual analytics approach to analyze movement patterns across cities through social media data. Blaas et al.~\cite{blaas_2009_smooth} proposed a smooth curved line to visually support exploring the transition between states in an observational time series at a macro scope, while it is not suitable for visualizing various higher-order dependencies exist on one physical node together. Rosvall et al.~\cite{rosvall2014memory} proposed a glyph that bridges the two consecutive nodes directly by the current node. In theory, this method can integrate different dependencies through the current node to build the correspondence, but the area of the current node limits the scalability of various dependencies. To handle such issues, HoNVis~\cite{tao2017honvis} visualizes the higher-order dependencies and supports the exploration from an overview to a fine-grained level while ignoring the temporal characteristics of the higher-order dependency. \holens~aims to intuitively represent the higher-order dependencies, especially revealing the temporal features that can further help experts analyze the movement pattern.


\section{Requirement and Method Overview}\label{sec: holens-Requirement Analysis and Method overview}


In this section, we summarize the requirements~(\sect\ref{sec holens-requirement}) and provide an overview~(\sect\ref{Sec: holens-overview}) of our solution.

\subsection{Requirement Analysis}\label{sec holens-requirement}

In the early stage of this research, we held regular meetings with the target users and two domain experts (\textbf{E.A} and \textbf{E.B}). The target users are focused on emergency response for traffic management and urban planning. They aim to find the law of human movement and determine what people tend to do (what happened) in a certain region at a certain time. Their daily work, which involves analyzing higher-order movement patterns, has two goals, including 1) segmenting the urban area into functional regions and 2) optimizing the traffic.
We summarize the requirements for this research from two aspects: \textbf{higher-order movement modeling} (R.1, R.2, R.3) and \textbf{higher-order pattern visualization} (R.4, R.5).


\vspace{1.5mm}
\noindent\textbf{R.1 Context-aware Aggregation.}~
\fzz{To avoid sparsity of the movement data in an urban area, aggregating the data into clusters and constructing a DAG based on these clusters is essential for analysis.} Furthermore, when aggregating the clusters, considering the urban contextual information (\eg, the function of the block and the integrity of the buildings) can facilitate the analysis of higher-order patterns in which more movements with similar features are gathered together.


\vspace{1.5mm}
\noindent\textbf{R.2 Multi-scale Self-organization.}~
Analysts usually explore higher-order patterns at different levels, such as district and city levels. Thus, the principle of merging the small regions into a bigger one and keeping the regions' consistency at different hierarchies profoundly influences the result of analyzing the higher-order pattern.

\vspace{1.5mm}
\noindent\textbf{R.3 Higher-order Pattern Extraction with Temporal Variability.}~
Human activities usually depict temporal variability. Some activities only exist in certain periods, and some may implicitly exhibit peak and nonpeak features. Analyzing the higher-order patterns can help understand the traveling behavior and inspire them to make proper decisions, such as urban planning. Therefore, a mining method for extracting these higher-order patterns with temporal constraints is helpful.


\vspace{0.5mm}
\textbf{R.3.1 Global Higher-order Pattern Extraction.}~
Factors, such as daily commuting and weekend rest, often reflect higher-order patterns, \eg, from housing estate to business district and then back on weekdays. Therefore, the global movement trend between different regions must be explored, and the higher-order patterns between different regions within the overall scope of the city must be mined.


\vspace{0.5mm}
\textbf{R.3.2 Local Higher-order Pattern Extraction.}~
In contrast to extracting higher-order patterns globally, local higher-order pattern extraction is performed for a given region. As a result, these patterns are more specific, and the temporal variability is more detailed. However, the amount of movement data in the given region may sometimes decrease dramatically due to small ranges. Therefore, the flow of higher-order patterns must also be considered to avoid overfitting during extraction.

\vspace{1.5mm}
\noindent\textbf{R.4 Higher-order Pattern Visualization.}~
Representing a higher-order pattern for a global region or a given local region is similar to visualization. A good visualization design for higher-order patterns can help summarize the travel mode. In addition, for one node in a DAG, its meaning in a higher-order pattern differs from its physical meaning at the DAG. Usually, one node in a DAG may represent different orders of many higher-order patterns. Therefore, the visualization of the nodes in the DAG and higher-order aspects is essential. 

\vspace{1.5mm}
\noindent\textbf{R.5 Higher-order Pattern Comparison.}~
For the higher-order patterns at different temporal and geographical dimensions, it is helpful to know the existence time of different higher-order patterns, their geographical distribution on the map, and the movement composition at each node.

\subsection{Method Overview}\label{Sec: holens-overview}
\holens~consists of two phases, including movement modeling (\sect\ref{sec: holens-model}) and interactive higher-order pattern visual exploration (\sect\ref{sec: holens-visualizations}). The movement modeling phase supports adaptive, hierarchical movement aggregation and higher-order pattern extraction. In the interactive visual exploration phase, a novel visual analytics interface is designed for interactive analysis.


\begin{figure}[!ht]
\centering
  \includegraphics[width=\linewidth]{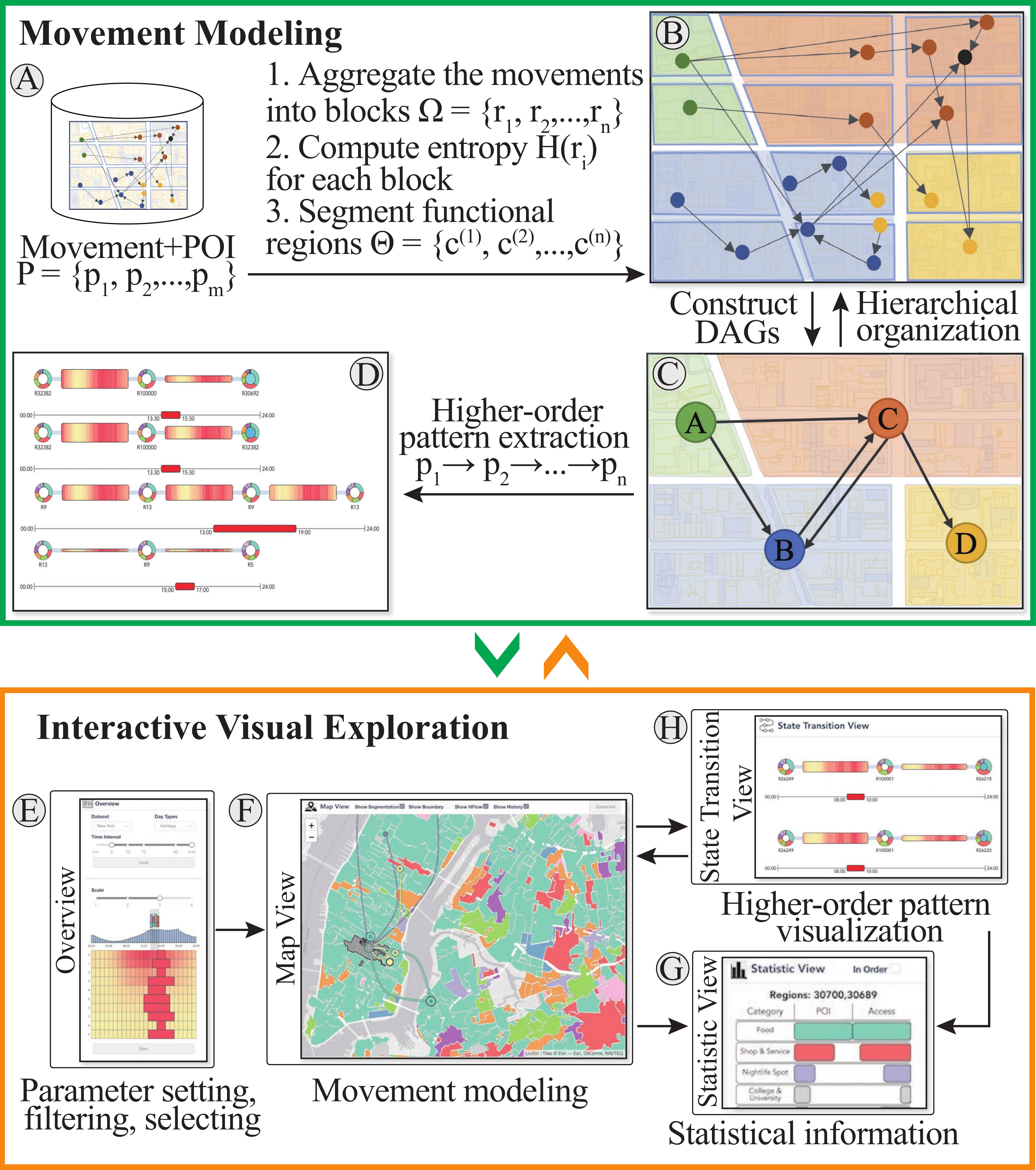}
  \vspace{-1.5em}
  
  \caption{The overview of \holens. \holens~consists of two stages: movement modeling (A, B, C, and D) and interactive visual exploration (E, F, G, and H).}
  
  \label{fig:holens-overview}
   \vspace{-1em}
\end{figure}

In the movement modeling phase, \holens~takes movement data and POI categories as inputs (\fig\ref{fig:holens-overview}($A$)). 
After cleaning the data, \holens~first estimates the stay time of each moving object in each location. Next, 
\holens~aggregates the movements in the ``regions” and distinguishes the region’s dominant functions.
Then, the basic ``regions” are self-organized according to their entropy by the proposed aggregation algorithm (\fig\ref{fig:holens-overview}($B$)). In addition, the regions can be organized at different scales. Afterward, a DAG is constructed with each region as a vertex and the flow between the regions as the edge (\fig\ref{fig:holens-overview}($C$)). In accordance with the DAG, the higher-order patterns are extracted, with consideration to temporal variability (\fig\ref{fig:holens-overview}($D$)).

For the interactive visual exploration phase, an interactive visual analytics interface (\fig\ref{fig:holens-teaser}) is developed. It is a web-based multi-view interface that aims to provide interactive visual analytics for higher-order pattern exploration. Data collection and preprocessing are run offline on a GPU server with four GeForce GTX 2080Ti graphics cards. \holens~runs on an Apple M1 MacBook Pro with 16 GB memory. The backend is supported by Flask, and the frontend is implemented by Vue.js and D3.js. In the overview (\fig\ref{fig:holens-overview}($E$)), users can observe the distribution of the global higher-order patterns in the temporal dimension. In the map view (\fig\ref{fig:holens-overview}($F$)), users can visualize the results of movement aggregation and the spatial distribution of the higher-order patterns. Finally, the state transition view (\fig\ref{fig:holens-overview}($H$)) visualizes the higher-order patterns in detail and compares different higher-order patterns. Moreover, the statistic view (\fig\ref{fig:holens-overview}($G$)) displays the statistics from both the map view and the state transition view.


\section{Model}\label{sec: holens-model}
This section first introduces the construction of a data structure (\sect\ref{sec: holens-data preparetion}). Next, a novel method for data aggregation (\sect\ref{sec: holens-data abstraction}) and hierarchical movement self-organization (\sect\ref{sec: holens-segmentation}) is introduced. Finally, the method for higher-order pattern extraction in this research (\sect\ref{sec: holens-higher-order dependency extraction}) is introduced. 

\subsection{Data Preparation}\label{sec: holens-data preparetion}
Urban movement data usually exhibit three essential attributes: the ID of the moving object, the position of the spatial dimension, and the timestamp of the temporal dimension. In map visualization, the geographical data from the open street map (OSM)\footnote{\href{https://www.openstreetmap.org/}{https://www.openstreetmap.org/}} is also utilized. Although each movement data record contains a timestamp that can reflect the time point of one moving object existing in a certain position, it cannot provide the period each moving object stays in the same position. Thus, in this research, a shortest path-based method is applied to estimate the period of one moving object's stay in a certain position. We first form the sequence of each unique moving object's position according to the corresponding timestamp in one day to generate their movement trajectory. Next, the trajectory with significant time intervals is divided into independent sub-trajectories, in which the two consecutive points with large intervals have little associations, and then the sub-trajectory with just a single point is filtered out. Afterward, the travel time of two adjacent movement points is computed in accordance with the traffic condition and the optimal path by using the mainstream commercial map application (\ie, the API from Google Maps\footnote{\href{https://www.google.com/maps/}{https://www.google.com/maps/}}). Finally, the moving object's stay time at each position is deduced by subtracting the travel time between two adjacent points.

\begin{table}[h]
    \centering
    \vspace{-1em}
    \caption{\fzz{Summary of Major Variables}}
    \begin{tabular}{|l|p{0.6\linewidth}|}
        \hline
        \textbf{Notation} & \textbf{Description} \\
        \hline
        $\Omega$ & Set of spatial regions \\
        \hline
        $p_j$ & $j$-th Point of Interest (POI) \\
        \hline
        $\Theta$ & Set of POI centroids \\
        \hline
        $D_i$ & Density vector for region $r_i$ \\
        \hline
        $d$ & Number of distinct POI categories \\
        \hline
        $\alpha$ & Aggregation threshold \\
        \hline
        $\mathrm{O}_i$ & Set of POIs projected into region $r_i$ \\
        \hline
        $\mathrm{H}(r_i)$ & Entropy of region $r_i$ \\
        \hline
        $g$ & R tree function mapping POIs to regions \\
        \hline
        $i^{k+1}$ & Vertex in the constructed DAG \\
        \hline
        $i_{\Delta{T}}^k$ & $k$-th vertex in the DAG for a specific time period $\Delta{T}$ \\
        \hline
        $\Delta{T}$ & Selected time period \\
        \hline
        $\omega{(i_{\Delta{T}}^k\rightarrow i^{k+1})}$ & Edge weight indicating connection intensity in period $\Delta{T}$ \\
        \hline
        $X^{k+1}$ & Vertex in the DAG at time step $k+1$ \\
        \hline
        $I$ & Sequence of vertices comprising the higher-order dependency \\
        \hline
    \end{tabular}
    \vspace{-1em}
    \label{tab:variables}
\end{table}

\subsection{Aggregation and Hierarchical Organization}\label{sec: holens-data abstraction and segmentation}
\whj{In this section, we delve into the procedure of data aggregation (\textbf{R.1}) and the hierarchical self-organization of movement data (\textbf{R.2}).  Our aim here is twofold: 1), to consolidate movement points based on their spatial proximity while considering urban contextual information, and 2), to facilitate the adaptive formation of regions in response to user demand.  Subsequently, we define the concept ``region" following the work from~\cite{yuan2012discovering}, which employed Connected Component Labeling (CCL) on a map divided by the road network. Moreover, Entropy can be used to identify higher-order patterns, as higher-order patterns are patterns that involve diverse POIs visited. For example, a higher-order pattern in human mobility data might be a sequence of visits to different types of POIs, such as a home, a workplace, and a restaurant. By combining entropy, we can easily achieve this. The scale of the region can be modified manually by selecting different types of roads (\eg, freeway, primary, secondary). Here, these regions serve as the fundamental building blocks for both data aggregation and point organization.}

\subsubsection{Data Aggregation}\label{sec: holens-data abstraction}
\whj{Our primary focus in this research is to gain insights into events occurring within Points of Interest (POIs) located within specific regions. To achieve this, we begin by mapping movement data, originally projected onto roads, to the nearest region.  In the context of data aggregation, entropy can be used to measure the uncertainty of the POIs within a region. A region with high entropy has a diverse set of POIs, while a region with low entropy has a more homogeneous set of data points.  Subsequently, we aggregate movement data within these corresponding regions and calculate the entropy of each region using the following steps:}

\whj{\noindent\textbf{Spatial Projection:} Formally, let $r_i$ denote the $i^{th}$ region obtained by CCL, and  $\forall r_i \in \Omega$  is the set of spatial regions. Then, we project each POI $p$ into the spatial region through the R tree~\cite{beckmann1990r} with 
$g=Rtree(\Omega)$, where $g:p \rightarrow r$ is the function of the R tree that returns the located region of $p$. Thus, the projected process for  region $r_i$ can be written as  
\begin{equation}
	\mathrm{O}_i=\left\{  c^{(j)} \ \mid \ g(p_j)=r_i   \right\}, \forall p_j \in \Theta,
\end{equation}
where $\Theta$ is the set of POI.  }

\whj{\noindent\textbf{ Entropy Calculation:}
Subsequently, we calculate the entropy for  region $r_i$,
\begin{equation}
	\mathrm{H}(r_i)=-\sum_{a_i \in D_i} P(a_{i}) \log _{e} P(a_{i}),
\end{equation}
where $D_i \in \mathbb{R}^{d}$ is the density vector calculated by $	\mathrm{O}_i$, where $d$ represents the number of distinct POI categories obtained from external sources, such as the Foursquare API\footnote{\href{https://api.foursquare.com/v2/venues/categories}{https://api.foursquare.com/v2/venues/categories/}}(\eg, \textit{Arts  \& Entertainment, Shop \& Service, and  Outdoors \& Recreation}). For example, $D_{i,j}$ indicates the density probability of the $j^{th}$ POI category in $r_i$, so that it satisfies $\sum_{j}D_{i,j}=1$. The entropy $H$ serves as the degree of confusion within the region since we aim to ignore the human mobilities that have many consistent attributions. For example, such higher-order dependencies: $Tr: p_1 \rightarrow p_2 \rightarrow \cdots \rightarrow p_n$, containing multiple repetitive POI types, are undesirable.  The process of this step is shown in ~\alg\ref{alg:holens-alg1}. }

\begin{algorithm}
	\caption{Region Entropy Computation}\label{alg:holens-alg1}
	\LinesNumbered
	\KwIn{spatial region set $\Omega=\{r_1,r_2,...,r_n\}$, set of GPS points $P=\{p_1,p_2,...,p_m\}$.}
    \KwOut{$\Theta_r=\{c^{(1)},c^{(2)},...,c^{(n)}\}$}
    $g=Rtree(\Omega)$  \quad // ST query\\
    \For{ $r_i \in \Omega$}{
        $\mathrm{O}_i=\left\{  c^{(j)} \ \mid \ g(p_j)=r_i   \right\}, \forall p_j \in P,$ \\
        $	\mathrm{H}(r_i)=-\sum_{a_i \in D_i} \mathrm{P}\left(a_{i}\right) \log _{e} \mathrm{P}\left(a_{i}\right)$
    }
    \For{ $r_i \in \Omega$}{
    	$\Theta \leftarrow \Theta + $ BFS($r_i$,\ $V$)
    }

	\textbf{return} $\Theta$
\end{algorithm}
\vspace{-1em}

\subsubsection{Hierarchical Movement Organization}\label{sec: holens-segmentation}

\whj{We introduce an innovative algorithm for hierarchical self-organization of movement data. This algorithm facilitates the clear visualization and exploration of higher-order dependencies at different levels while ensuring the consistency of regions across these levels. Essentially, it refines the spatial layout, reducing spatial entropy as mentioned in \sect\ref{sec: holens-data abstraction}.  Technologically, we concatenate those regions that present lower entropy after aggregating. Given the set $\Omega=\{r_1,r_2,...,r_n\}$, Breadth-First Search (BFS) is employed to search the associated neighbors for each region. The aggregated condition for every pair of regions is defined as }
\begin{equation}
	\frac{1}{2}(\mathrm{H}(r_i)+\mathrm{H}(r_j)) \geq \alpha\mathrm{H}(r_i+r_j),
\end{equation}
where $\alpha$ indicates the aggregating threshold. Intuitively, the smaller $\alpha$, the easier the aggregation. \alg\ref{alg:holens-alg2} gives the pseudocode for generating spatial proximity, returning $\Theta=\{c^{(1)},c^{(2)},...,c^{(n)}\}$, the set of renewed spatial regions. $c^{(i)}$ represents the cluster index corresponding to region $r_i$. 

As in \textbf{R.2}, \holens~aims to provide the multi-level region scale for systemically understanding the higher-order movement pattern. The first-order merged procedure has been introduced above; thus, for the next-level aggregation, only those regions incorporated in the previous iteration are labeled as new regions, and their inner regions are removed accordingly. This strategy is consistent with different level region scales because the high-level region styles always come from the lower-level regions. 

\begin{algorithm}
	\caption{BFS}\label{alg:holens-alg2}
	\LinesNumbered
	\KwIn{spatial region set $\Omega=\{r_1,r_2,...,r_n\}$, entropy $H$, visited set $V$, $\Theta_r=\{\}$.}
	\KwOut{$\Theta_r=\{c^{(i)}, c^{(j)}, ... \}$}
    \If{$r_i \notin V$}{
	\For{$r_j \in r_i.Neighbors $}{
		\If{$\frac{1}{2}(\mathrm{H}(r_i)+\mathrm{H}(r_j)) \geq \alpha\mathrm{H}(r_i+r_j)$}{
		$r_i \leftarrow merge(r_i,r_j)$\\
		 $r_i.Neighbors \leftarrow r_j.Neighbors \ \cup \ r_i.Neighbors $\\
		 $V \leftarrow V + r_j$\\
		 $\Theta_r \leftarrow \Theta_r + c^{(j)}$
	}
	}
	\For{$r_j \in r_i.Neighbors $}{
	 BFS($r_i,V$)
	}}
	   
\textbf{return} $\Theta_r$
\end{algorithm}
\vspace{-1em}
\subsection{Higher-order Dependency Extraction}\label{sec: holens-higher-order dependency extraction}

After aggregating the movements into regions, we construct a DAG with the centroid of each region as the vertex and the in/outflow between regions as the directed edges to extract the higher-order patterns (as illustrated in \fig\ref{fig:holens-overview}($C$) and \fig\ref{fig:holens-overview}($D$)). In this study, we consider the temporal variability in the higher-order patterns. Thus, we first modify the transition probability for the first-order Markov model by adding a temporal constraint~(\textbf{R.3}): 

\begin{equation}
P(X^{k+1} = i^{k+1}|X_T^k = i_{\Delta{T}}^k) = \frac{\omega{(i_{\Delta{T}}^k\rightarrow i^{k+1})}}{\sum_{j}\omega{(i_{\Delta{T}}^k\rightarrow j)}},
\label{equ:holens-first-order}    
\end{equation}
where $i_{\Delta{T}}^k$ indicates the $k$th vertex in the constructed DAG, and $\Delta{T}$ denotes the selected period. The edge weight $\omega{(i_{\Delta{T}}^k\rightarrow i^{k+1})}$ indicates the intensity of the connection in a certain period $\Delta{T}$, which in our scenario can be assigned as the sum of pairwise connections $(i^k\rightarrow i^{k+1})$ in period $\Delta{T}$.

For higher-order pattern extraction, many existing methods~\cite{schwarz1978estimating, akaike1974new, chierichetti2012web, van1998testing} focus on a fixed order, but we focus on the number of orders that is not fixed; thus, we refer to the method of Xu et al.~\cite{xu2016representing} and modify the method by adding temporal constraints.
\begin{small}
  \begin{equation}
P(X^{k+1} = i^{k+1}|X_T^k = (i_{\Delta{T}}^k | I)) = \frac{\omega{((i_{\Delta{T}}^k|I)\rightarrow i^{k+1})}}{\sum_{j}\omega{((i_{\Delta{T}}^k|I)\rightarrow j)}}, 
\label{equ:holens-higher-order}    
\end{equation}  
\end{small}
where $I$ indicates a sequence of vertices comprising the higher-order dependency, \eg, if a higher-order dependency exists at vertex D, that is, $[D|C,B,A]$. $I$ represents the sequence $C,B,A$, and $i_{\Delta{T}}^k | I$ indicates that $D$ with the higher-order dependency $[D|C,B,A]$ exists in $\Delta{T}$. 

We take advantage of the Kullback-Leibler divergence (KLD)~\cite{kullback1951information} to measure recursively whether we should enlarge the current order, in other words, to control the order number of the higher-order pattern. \fzz{For example, we measure the transition probability of the current $n$-order dependency and its ($n+1$)-order dependency by KLD; if it significantly changes, it indicates that the dependency should be increased from the current order to ($n+1$)-order.} 


\begin{figure*}[ht]
  \centering
  \includegraphics[width=\textwidth]{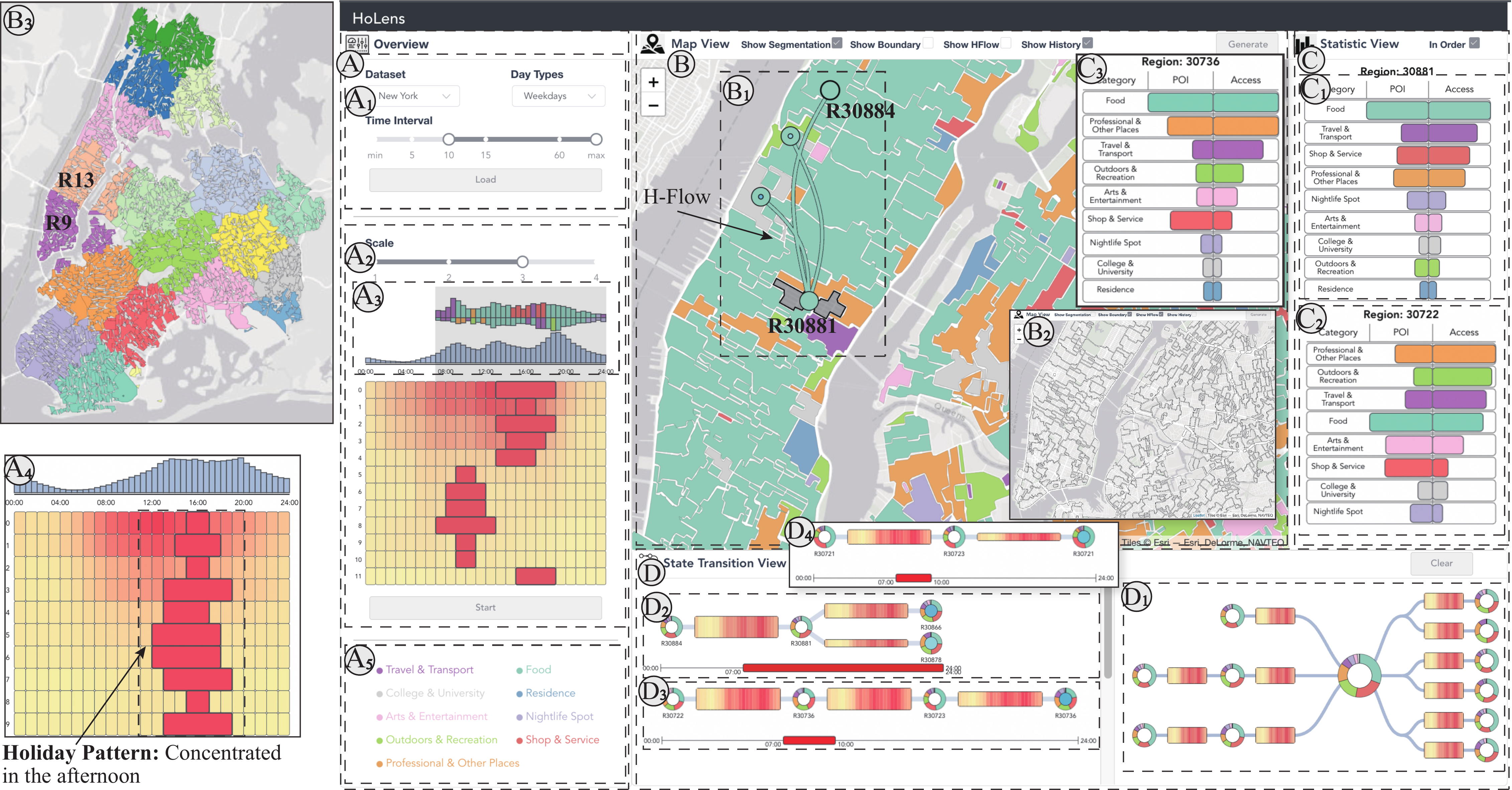}
  \caption{\holens~supports visually exploring the urban higher-order patterns, which consists of four views: (A) the overview provides a configuration panel and a global representation to the end-users, (B) the map view displays the result of movement aggregation and supports visualizing the higher-order patterns in a spatial dimension, (C) the statistic view shows the detailed statistical information of the selected region, and (D) the state transition view provides the function for the further exploration the higher-order patterns.}
  \label{fig:holens-teaser}
  \vspace{-1.5em}
\end{figure*}


\section{Visualizations}\label{sec: holens-visualizations}
This section introduces the designs of \holens, which helps the analysts interactively analyze the higher-order patterns.

\subsection{Overview}\label{sec:holens-visualization-overview}

The overview aims to help analysts configure the parameters and control the analytical process from the temporal dimension. The analysts can select the dataset, pre-configure the parameters, control the temporal process, and obtain the temporal distribution of the higher-order patterns. The overview consists of a configuration panel~(\fig\ref{fig:holens-teaser}($A_1$)), a global controlled temporal distribution matrix (\fig\ref{fig:holens-teaser}($A_2$)), and a legend board(\fig\ref{fig:holens-teaser} ($A_5$)).

The configuration panel (\fig\ref{fig:holens-teaser}($A_1$)) supports dataset selection, analytical day types selection (weekday/weekend), and time interval filter. 
The global controlled temporal distribution matrix consists of a controlled timeline (\fig\ref{fig:holens-teaser}($A_3$)) and a temporal distribution matrix (\fig\ref{fig:holens-teaser}($A_4$)) (\textbf{R.4}). The controlled timeline supports controlling the period and visualizing the flow from the temporal dimension (\textbf{R.3}). It~(\fig\ref{fig:holens-teaser}($A_3$)) is composed of a time axis, a single-directional histogram, a bi-directional histogram, and a sliding window. 
The single-directional histogram reflects the tendency of the active movements distributed in one day. 
The sliding window aims to enhance the interaction when analyzing the detailed higher-order patterns. When the analysts select an AOI in the map view~\fig\ref{fig:holens-teaser}($B_1$), the bi-directional histogram \fig\ref{fig:holens-teaser}($B_3$) is generated to visualize the in-/out-flow of the selected region. The color scheme of the bar corresponds to the POI category, indicating the function of the selected region in this period, and the height of the bi-directional histogram reflects the volume of the flow in detail. 

The temporal distribution matrix~(\fig\ref{fig:holens-teaser}($A_4$)) reflects the global higher-order patterns distribution at a temporal dimension. It jointly shares the same time axis with the controlled timeline. 
Each row of the matrix is a heatmap that reflects the flow of the higher-order dependencies. In each row, the colored bar is located at the corresponding position to visualize the city-scale higher-order pattern and is placed in descending order of flow size. When the analysts select one higher-order pattern in the temporal distribution matrix, a higher-order state sequence chart~(\fig\ref{fig: holens-case study2 global higher-order pattern}($B$)) is generated (\fig\ref{fig:holens-teaser}($D$)). 

\subsection{Map View}\label{sec: holens-mapview}

The map view~(\fig\ref{fig:holens-teaser}($B$)) is designed to support exploring the higher-order patterns in a spatial dimension. Analysts can interact with other views (overview and state transition view) through the map view, and the result in spatial dimension can be displayed in the map view.

\vspace{1.5mm}
\noindent\textbf{Region segmentation.} The map view supports segmenting the urban areas, providing an intuitive representation of the movement aggregation on the map. As mentioned in~\sect\ref{sec: holens-data abstraction and segmentation}, ``region" is used as the basic unit; thus, the segmented region is composed of these basic ``regions."
In the map view, all the single unit ``regions" are aggregated in the same segmented region, and the in-between roads and boundaries are eliminated for a more intuitive visual effect. The region segmentation function in the map view reveals the dominant POI category for the selected period (selected in the controlled timeline~(\fig\ref{fig:holens-teaser}($A_3$)) in the overview) at each segmented region, enabling the analysts to understand efficiently what people in the segmented region tend to do. 

\vspace{1.5mm}
\noindent\textbf{Comparison \& Multi-region Exploration.}~The map view supports comparison~(\textbf{R.4}) and multi-region exploration, helping analysts compare the movement aggregation at different periods and explore higher-order patterns based on the user demand. The comparison function in the map view is different from that in the state transition view (\fig\ref{fig:holens-teaser}($D$)). In the map view, the comparison is used to compare the aggregation results at different periods. In \sect\ref{sec: holens-data abstraction}, the method enables adaptive aggregation, and the aggregation result changes over time. Therefore, a grid design~(\fig\ref{fig:holens-case1}($D$) to ($G$)) is used to compare the historical selection. When analysts adjust the sliding window~(\fig\ref{fig:holens-teaser}($A_3$)). The movement aggregation result changes in the map view, but the previously selected region remains~(\fig\ref{fig:holens-case1}($D$) to ($G$)) for comparison in spatial dimensions. When analyzing the higher-order pattern, analysts sometimes need to analyze a continuous group of regions. Therefore, users can click one or a group of regions to analyze, and \holens~may see these regions as a whole to extract the higher-order patterns.

\begin{figure}[h!]
\centering
  \includegraphics[width=\linewidth]{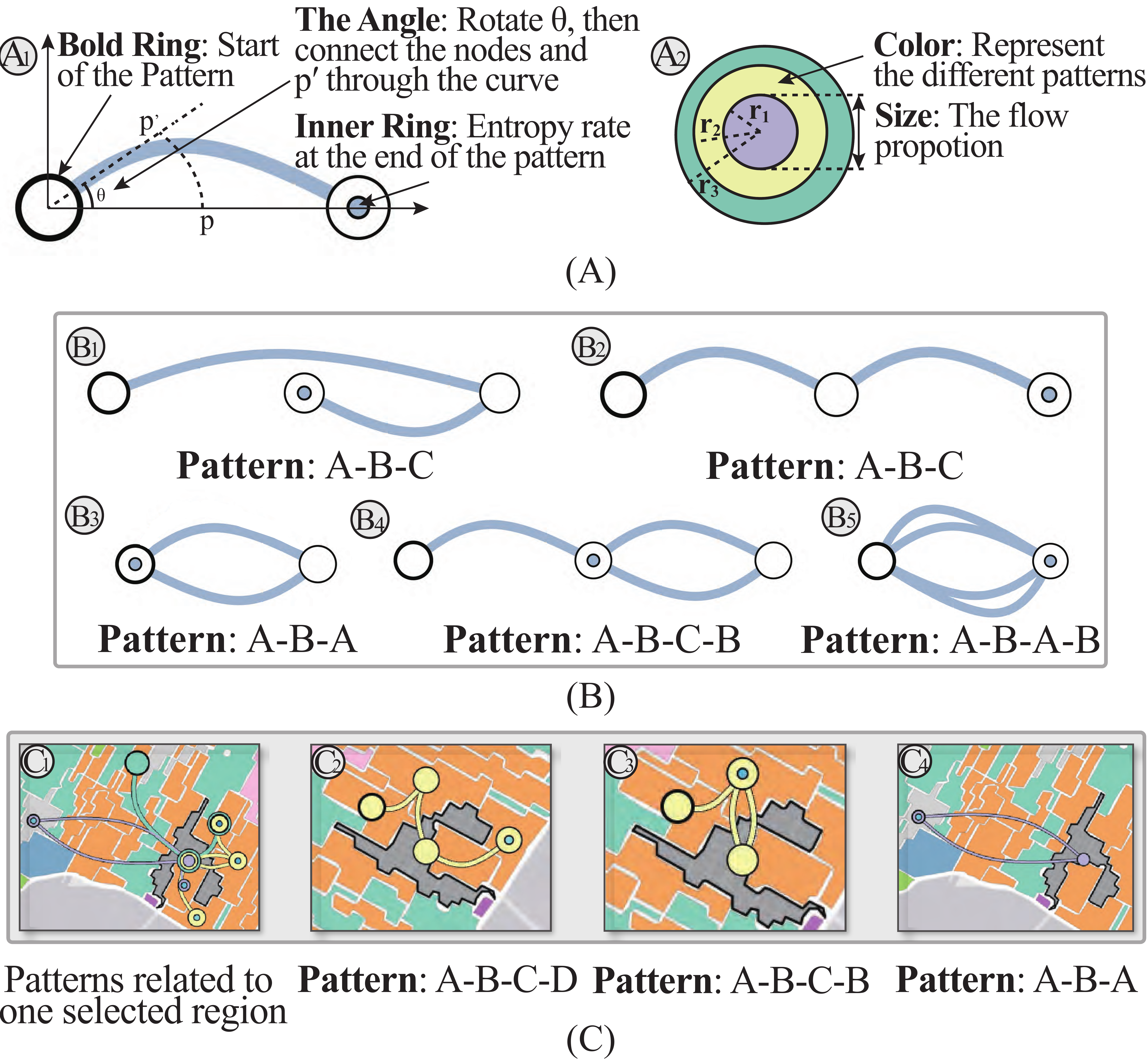}
   \vspace{-2em}
    \caption{Illustration of H-Flow. (A) The design scheme of H-Flow, which aims to represent the higher-order dependencies in (B) the higher-order pattern from the (C) spatial dimension.}
    \label{fig: holens-hflow}
  \vspace{-0.5em}
\end{figure}

\vspace{1.5mm}
\noindent\textbf{Higher-order Pattern Visualization.}~Inspired by Kelpfusion~\cite{meulemans2013kelpfusion}, a novel designed flow visualization (\fig\ref{fig:holens-teaser}($B_1$)) named \textbf{H}igher-order \textbf{Flow} (H-Flow) is proposed to visualize the higher-order dependencies from the spatial dimension~(\textbf{R.5}). The details of the H-Flow are as follows:

\noindent\textit{Construction Scheme} (\fig\ref{fig: holens-hflow}($A$)): The first node of H-Flow is designed as a bold circle, representing the beginning of the higher-order pattern~(\fig\ref{fig: holens-hflow}($A_1$)). For two adjacent orders in the higher-order dependency, they are linked by a curve~(\fig\ref{fig: holens-hflow}($A_1$)). First, we rotate the midpoint $p$ at an angle $\theta$, and then connect the two nodes and the rotated point ($p'$) sequentially controlled by the Bezier curve function~\cite{baydas2019defining}. We regulate a deflection in the same direction (clockwise/counterclockwise) to represent the direction. 
As~\cite{tao2017honvis}, we provide a representation of the entropy rates~\cite{rosvall2014memory,xu2016representing} of each extracted higher-order pattern, indicating the certainty, that is, the more certain, the lower entropy rates. To identify the end state of the pattern, we nested a small circle~(\fig\ref{fig: holens-hflow}($A_1$)) in the last state of the pattern, that is, the color saturation, to indicate the certainty of the whole pattern. Additionally, to avoid the loss of information caused by stacking at the junction when an intersection exists, we use the radius to indicate the proportion of the flow, which is the larger flow on the lower layer and distinguished by different colors~(\fig\ref{fig: holens-hflow}($A_2$)). 

\noindent\textit{Details of the H-Flow} (\fig\ref{fig: holens-hflow}($B$)):~This research focuses on exploring second-/third-order patterns (more discussion in~\sect\ref{sec: holens-order number}). Therefore, according to the real cases (\sect\ref{sec: holens-case-1}) and the feedback from the experts' interview (\sect\ref{sec: holens-expert-interview}), we summarized several higher-order pattern modes to regulate the design principles for H-Flow. The first mode is the linear mode, \eg, pattern $A-B-C$, pattern $A-B-C-D$. The second is the annular mode, in which the higher-order contains a cyclicity, and the destination of the cyclicity is the end of the pattern, \eg, pattern $A-B-A$. Here, we obey the construction scheme for these two situations; examples can be seen in~\fig\ref{fig: holens-hflow}($B_1$, $B_2$ and $B_3$). However, for other situations, \eg, patterns $A-B-A-C$ and $A-B-A-B$, the destination of the cyclicity is not the end of the pattern. Thus, to distinguish different flows emitted from the same node with the same direction, the second rotation at an angle of 2/3$\theta$ is made. Thus, it can distinguish the different flows in the higher-order patterns, \eg,~\fig\ref{fig: holens-hflow}($B_4$ and $B_5$). Furthermore, 
the corresponding higher-order patterns are generated in the state transition view~(\fig\ref{fig:holens-teaser}($D$)) in the form of the higher-order state sequence chart~(\fig\ref{fig:holens-teaser}($D_2$)) when clicking the node of the H-Flow. Moreover, to avoid interference from multiple colors that represent the different POI categories, we added the function of ``show boundary." The map view retains the boundary of the segmentation region~(\fig\ref{fig:holens-teaser}($B_2$)). 


\subsection{Statistic View}\label{sec: holens-statistic view}
The statistic view aims to show the detailed statistical information of the selected region intuitively~(\textbf{R.5}).
In \holens, we use tornado diagrams~(\fig\ref{fig:holens-teaser}($C_1$))
to represent the proportion of the POI categories in the selected region. 
Given that the number of POIs and the access frequency are of different orders of magnitude, these two parts are normalized to make the comparison more intuitive. Analysts can sort according to different demands (POI descending or access descending) for further analysis. Furthermore, in the statistic view, two tornado diagrams represent the statistical information from different views to provide an intuitive visual comparison. The upper (\fig\ref{fig:holens-teaser}($C_1$)) part displays the statistical information of the selected region in the map view. By contrast, the lower (\fig\ref{fig:holens-teaser}($C_2$)) part displays the information of the selected circular glyph in the state transition view, which is convenient for analysts to compare different regions in one pattern.

\subsection{State Transition View}\label{sec: holens-state transition view}
\begin{figure}[ht]
\centering
  \includegraphics[width=0.95\linewidth]{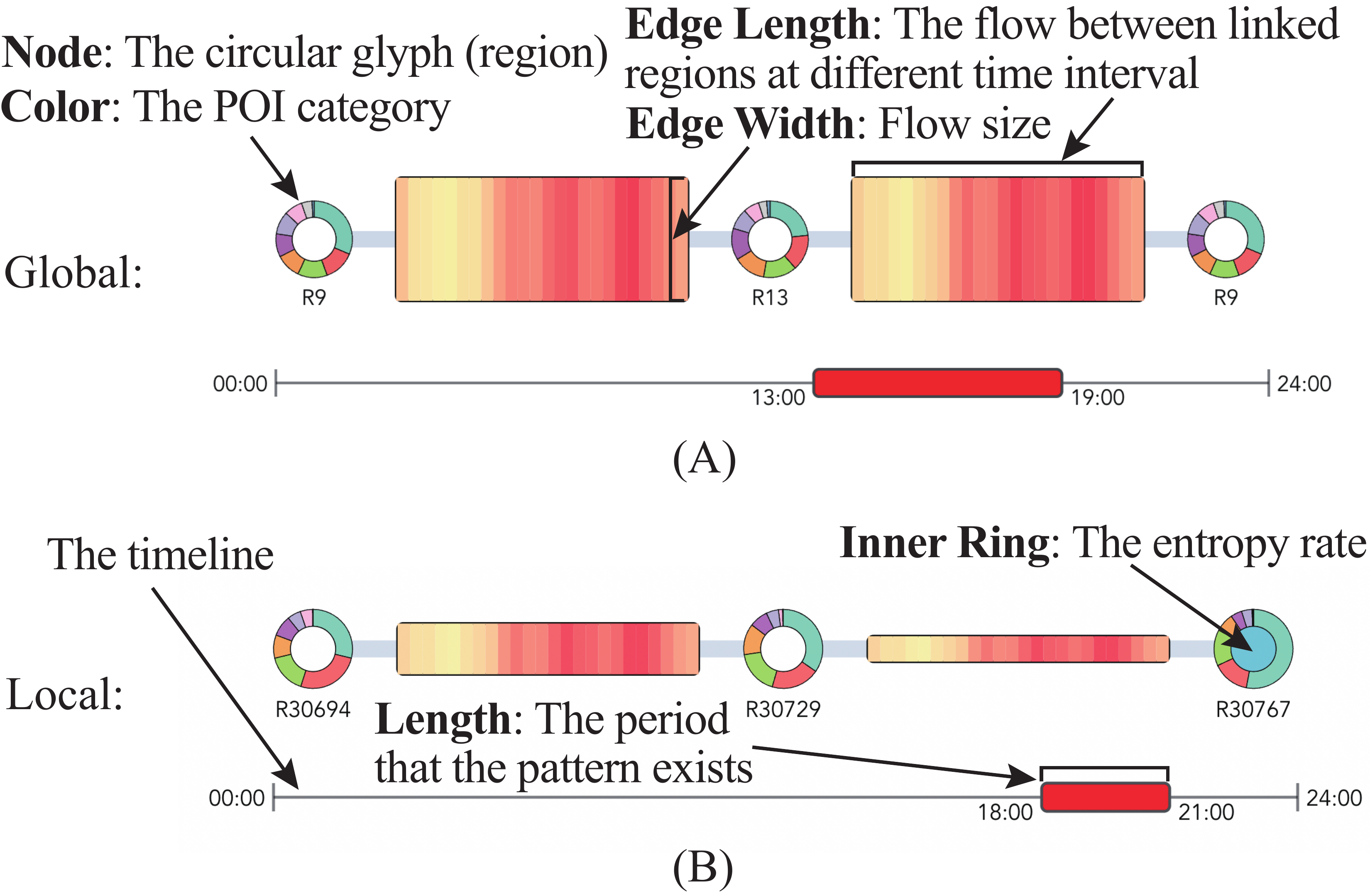}
    \caption{Illustration of the higher-order state sequence chart. The higher-order state sequence chart includes two types: (A) the global and (B) the local higher-order state sequence chart.}
    \label{fig: holens-state transition}
  \vspace{-1em}
\end{figure}

The state transition view~\fig\ref{fig:holens-teaser}($D$) aims to show the details of the higher-order patterns~(\textbf{R.4}) and support comparison between different higher-order patterns~(\textbf{R.5}). In this view, we design a higher-order state sequence chart~(\fig\ref{fig: holens-state transition}) that can not only represent the higher-order state but also reveal the flows' change between two states.

The higher-order state sequence chart is composed of nodes and edges. The node is a circular glyph~(\fig\ref{fig: holens-state transition}($A_1$)), and each node represents one region.
The sector of the circular glyph is the distribution of the flow for each category of POI. The edge~(\fig\ref{fig: holens-state transition}($A_1$)) between two adjacent nodes represents the flow between two regions. In comparison with the conventional Sankey diagram\cite{schmidt_2008_sankey}, 
we first use a rectangle~(\fig\ref{fig: holens-state transition}($A_1$)) located at the center between two adjacent nodes instead of the edge of the Sankey diagram to represent the flow. The width of the rectangle indicates the overall flow between two adjacent nodes. In accordance with Zeng et al.~\cite{zeng2016visualizing}, a heatmap is applied to represent the flow change between two regions in each time interval. 
It can also reveal the difference between first-order dependence and higher-order patterns at the temporal dimension. Given the temporal feature of the higher-order pattern, we add a timeline~(\fig\ref{fig: holens-state transition}($A_2$)) under the higher-order state sequence chart to represent the temporal dimension that the current higher-order pattern occurs. The design of the two kinds of patterns in the state transition view is similar. The only difference is that at the end of the local higher-order pattern, we add a colored dot~(\fig\ref{fig: holens-state transition}($A_2$)) to show the entropy rate of the local higher-order pattern, which is consistent with the design of H-Flow (\fig\ref{fig:holens-teaser}($B_1$)).

Specifically, H-Flow shows all local higher-order patterns related to the selected regions as a collection of higher-order patterns.
Correspondingly, the collection of the higher-order state sequence chart~(\fig\ref{fig:holens-teaser}($D_1$)) is generated in the right of the state transition view when the collection of the local higher-order patterns related to the selected region is visualized on the map. Analysts can select the higher-order pattern they want to explore by clicking the corresponding circular glyph in this collection. The higher-oerder sequence chart can be generated in two ways. The first one is to click the corresponding period of the higher-order patterns extracted in the grid matrix in the overview. The other is to click the corresponding node of H-Flow in the map view for further exploration. 
The state transition view also supports the comparison mode between different higher-order patterns. A higher-order state sequence chart is generated in the state transition view whenever the analyst selects a higher-order pattern. Analysts can visualize multiple higher-order patterns simultaneously to compare their flow distribution, the POI composition of the regions participating in the higher-order pattern, the formation time of the higher-order pattern, and the temporal distribution of the flow in two adjacent regions.
\begin{figure}[htbp]
\centering
  \includegraphics[width=0.9\linewidth]{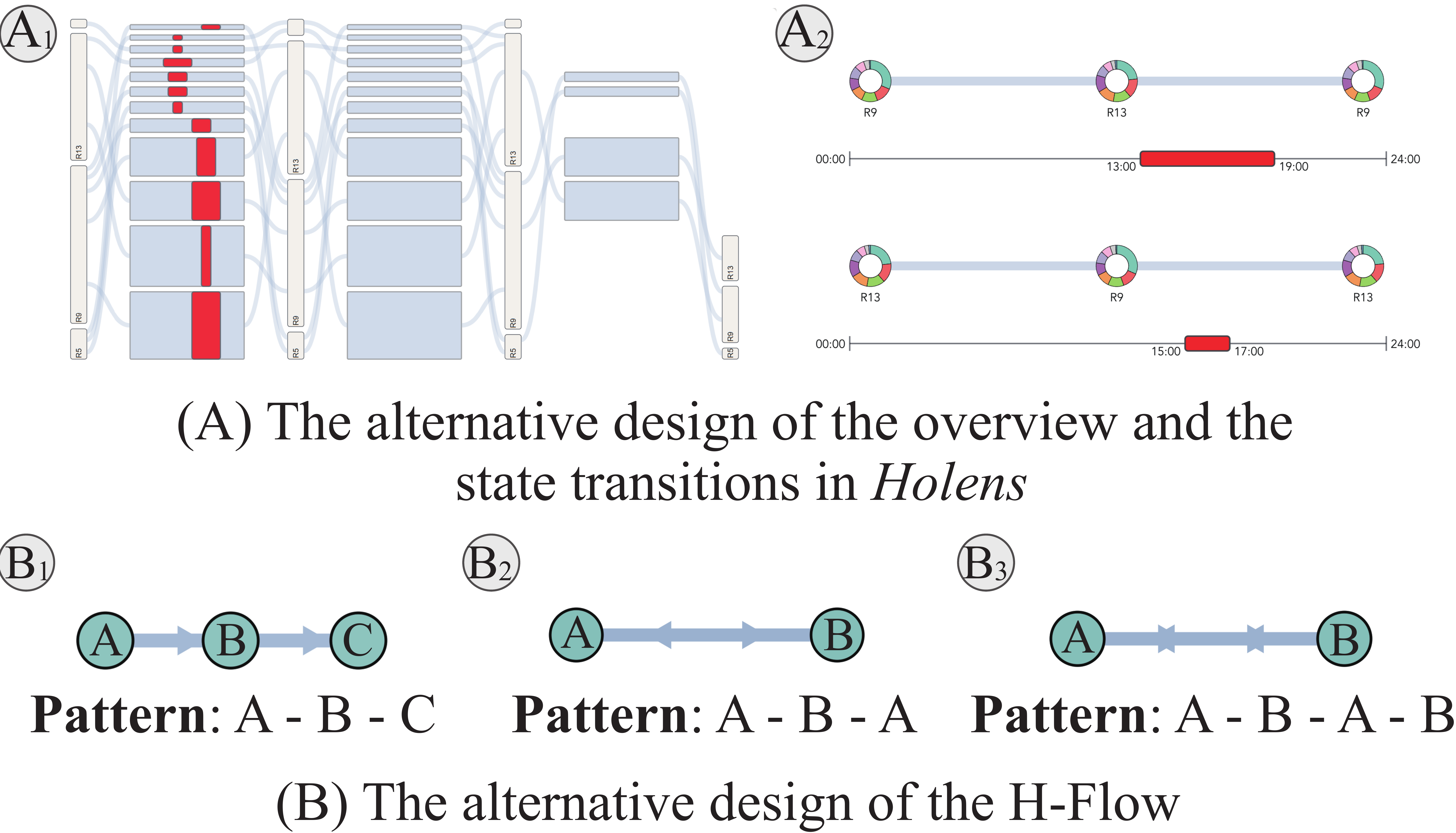}
  \vspace{-0.5em}
    \caption{The alternative designs. (A) is the overview and the global higher-order state sequence chart to visualize the higher-order patterns. (B) is the initial design of the H-Flow.}
    \label{fig: holens-alternative design}
  \vspace{-0.5em}
\end{figure}
\subsection{Alternative Designs}\label{sec: holens-alternative design}
\begin{figure}[htbp]
\centering
  \includegraphics[width=\linewidth]{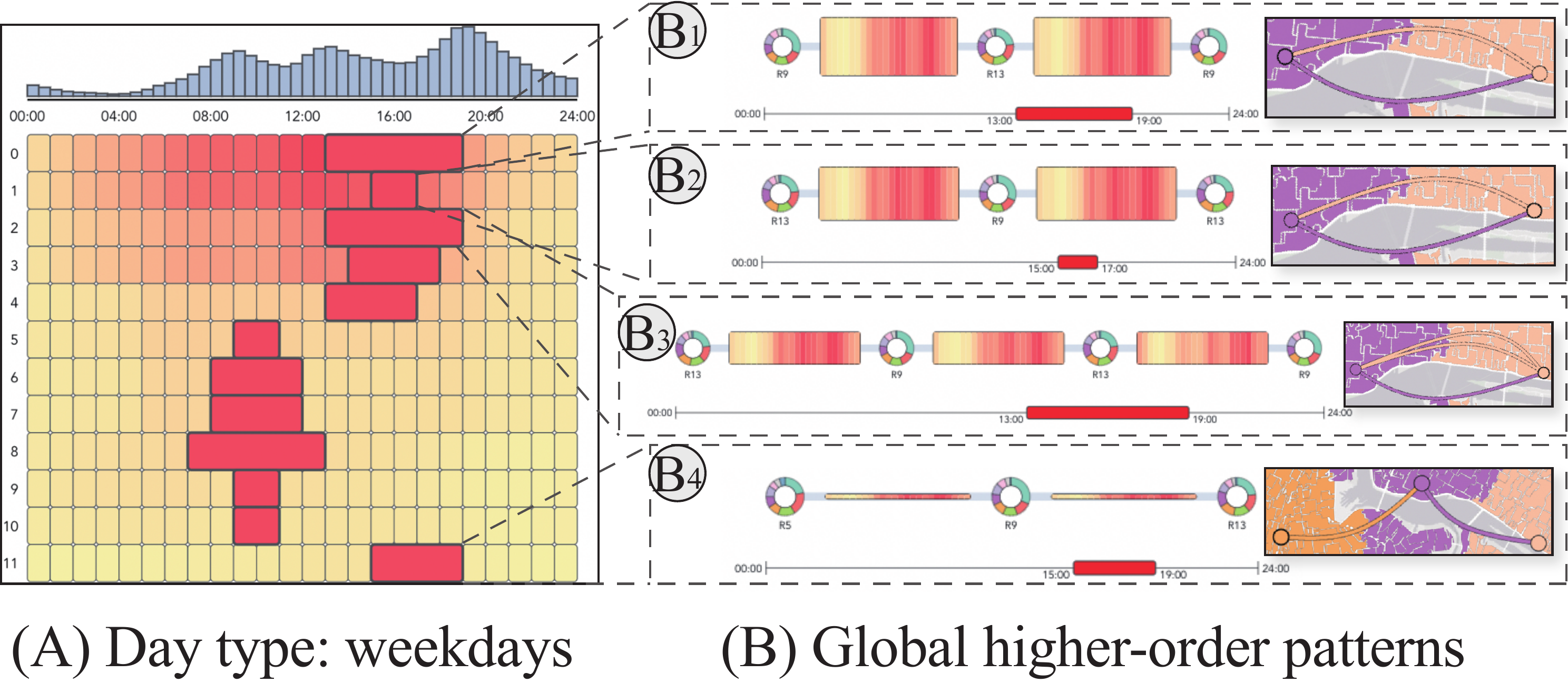}
    \vspace{-1.5em}
    \caption{From the exploration of global higher-order patterns in (A) the global controlled temporal distribution matrix, we summarize two modes: the the annular higher-
order pattern (\eg, $B_1$, $B_2$, and $B_3$)) and the linear higher-order pattern (\eg, $B_4$).}
    \label{fig: holens-case study2 global higher-order pattern}
    \vspace{-1em}
\end{figure}

We modify and iterate the design by closely communicating with the domain experts. 
\fig\ref{fig: holens-alternative design}($A$) is the first attempt for visualizing the global higher-order pattern. We referred to the design of the conventional Sankey diagram~\cite{schmidt_2008_sankey} and modified it to satisfy our demand. We aggregate the same regions at the same order level, visualized as one node. A rectangle located at the center between two adjacent nodes is used to represent the flow. At the first edge, the inside red rectangle represents the period of the pattern. \fig\ref{fig: holens-alternative design}($A_2$) shows the design of local higher-order patterns, in which a circular glyph is used to show the detail of the POI composition in each region.
However, the global and local higher-order patterns are essentially state transitions. That is, to move from one state (region) to another with sequence and temporal variability. Therefore, this design may be redundant, and many higher-order patterns will generate visual clusters due to the intersection of links. 
Thus, we choose the current design in \fig\ref{fig: holens-case study2 global higher-order pattern}($A$) as the final design.

The initial design~(\fig\ref{fig: holens-alternative design}($B$)) for the H-Flow is to link the adjacent nodes directly through the straight line and place an arrow one-third from the destination to indicate flow direction. However, during the interview with the domain experts, we found that the higher-order pattern sometimes may probably exist in the form of a cyclicity, \eg, $A-B-A$, $A-B-A-B$ or $A-B-A-C$, in which the origin and the destination of the higher-order pattern are the same. Therefore, when visualizing the patterns such as $A-B-A$, the two directions of the H-Flow will coincide, and too many arrows may bring visual clutter. To this end, the curve is used instead of a straight line to represent the flow between adjacent nodes~(\fig\ref{fig: holens-hflow}). The details of the final version design of the H-Flow are detailed in \sect\ref{sec: holens-mapview}.

\subsection{Interaction}
The interaction of \holens~can be summarized as follows:

\vspace{1.5mm}
\noindent\textbf{Filtering and modifying to support the exploration details-on-demand:}~Users can filter and modify the data, enabling them to focus the information according to their interests. For example, they can use a bi-directional slider to filter the period when people stay in the POI and modify the scale of the clustered movements on the map. \holens~supports exploration information on demands at different levels, at different granularities, and in different regions. Users can select and operate in every view, hover the AOI to explore the details, and brush the sliding window to filter according to their demand. \holens~also supports the simultaneous exploration of global and local higher-order patterns. 

\vspace{1.5mm}
\noindent\textbf{Linking and highlighting visualizations:}~In \holens, views are linked to each other. For example, when an analyst hovers the circular glyph of the higher-order state sequence chart in the state transition view, the corresponding region in the map view is highlighted automatically, which facilitates the user in understanding the spatial information.


\begin{figure*}[htbp]
\centering
  \includegraphics[width=0.9\linewidth]{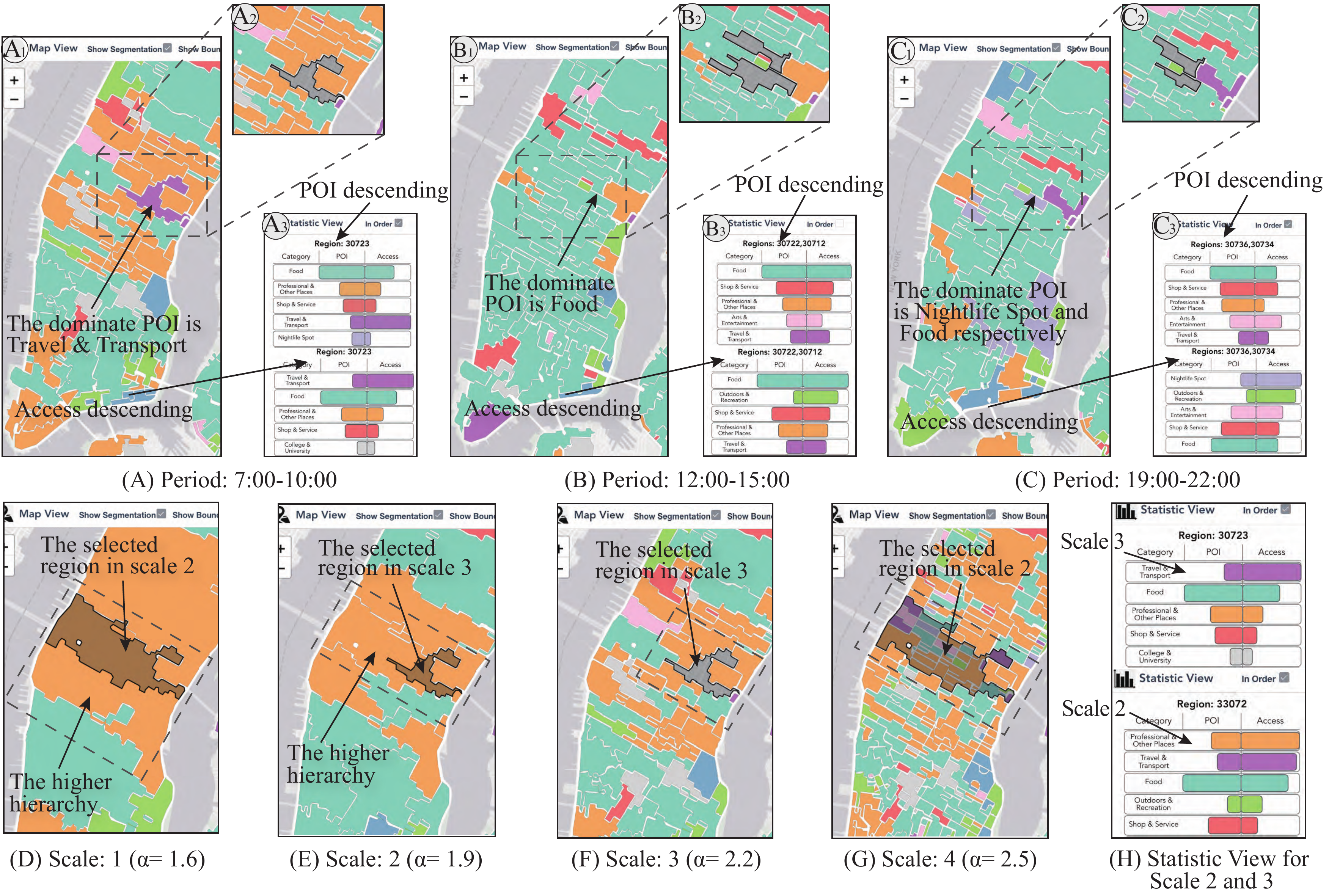}
  
  \caption{The movement modeling algorithm supports adaptively organizing the movement data with temporal variability (\eg, (A)~7:00-10:00, (B)~12:00-15:00, and (C)~19:00-22:00). Moreover, the algorithm also supports hierarchical multi-scale modeling (\eg, (D)~scale:1, (E)~scale: 2, (F)~scale: 3, and (G)~scale: 4), which the lower-level inherit the higher one.}
  \label{fig:holens-case1}
  \vspace{-1em}
\end{figure*}
\section{Evaluation}\label{sec: holens-evaluation}
This section conducts the case studies~(\sect\ref{sec: holens-case-1}, and \sect\ref{sec: holens-usage scenario}) manipulated by one of the collaborators (\textbf{E.A}) and the feedback from the interview of domain experts. The real-world check-in dataset~\cite{yang2019revisiting,yang2020lbsn2vec++} in New York City consists of 978054 check-in records from 3rd, Apr. 2012 to 29th, Jan. 2014.

\subsection{Study 1: Adaptive Organization and Hierarchical Multi-scale Aggregation}\label{sec: holens-case-1}
This case aims to analyze how \holens~aggregates urban movements. \textbf{E.A} first set the weekdays as the analytical day type. Then, he filtered the stay period for every movement as a default setting. After clicking the loading button, he found that the extracted higher-order patterns for the global scale are concentrated in the morning and the evening from (\fig\ref{fig: holens-case study2 global higher-order pattern}($A$)). In addition, except for these two periods, he also found from (\fig\ref{fig:holens-teaser}($A_3$)) that there are many active people at noon. Moreover, he found that the higher-order patterns are mostly concentrated in \textbf{R9} and \textbf{R13}, which correspond to midtown and downtown Manhattan. In this case, he selected \textbf{R9} and \textbf{R13} as the Area of Interest (AOI) for exploration, and the analytical periods were morning, noon, and evening.


\vspace{1.5mm}
\noindent\textbf{Adaptive Organization.}~First, \textbf{E.A} slide the sliding window to the morning (7:00-10:00 on the weekdays). The result of the movement aggregation is shown in \fig\ref{fig:holens-case1}($A_1$). He then arbitrarily selected one region (\fig\ref{fig:holens-case1}($A_2$)) for further exploration. The movement modeling and region segmentation results are shown in \fig\ref{fig:holens-case1}(A). By glancing at the statistic view (\fig\ref{fig:holens-case1}($A_3$)), \textbf{E.A} found that the distribution of POI shows that the dominant POI in this region is \textit{Food}, whereas the results by considering the access frequency are classified in the category \textit{Travel \& Transport}. \textbf{E.A} commented that the set period (7:00-10:00 on weekdays), in which people usually commute to work at this time and check in at bus stops/metro stations leads to the access frequency of \textit{Travel \& Transport} being much higher than any other POI category. Next, he moved the sliding window to another period (12:00-15:00). The movement aggregation and region segmentation results are shown in (\fig\ref{fig:holens-case1}($B_1$)). He found that the result for segmented regions changed, and the regions he previously selected (\fig\ref{fig:holens-case1}($A_2$)) are no longer the same. The category of this newly selected region is \textit{Food} (\fig\ref{fig:holens-case1}($B_3$)) by considering the weekday noon, which indicates that people in this region tend to have some food for lunch. Moreover, \textbf{E.A} then set the sliding window to the period at night (19:00-22:00). The movement modeling results are shown in (\fig\ref{fig:holens-case1}($C_1$)). The result in the related region (\fig\ref{fig:holens-case1}($C_2$)) is categorized as \textit{Nightlife Spot} (\fig\ref{fig:holens-case1}($C_3$)). \textbf{E.A} explained that according to his living experience in New York, this situation occurs because there are also many activities happening in the nightlife spots in this area after work.

\vspace{1.5mm}
\noindent\textbf{Hierarchical Multi-scale Movement Aggregation.}~In this exploration, \textbf{E.A} first changed the scale of the movement aggregation through a switch button in the map view. The movement aggregation result when he fixed the analytical period by the sliding window (07:00-10:00) and set one scale (scale 3) is shown in~\fig\ref{fig:holens-case1}($F$). Then, \textbf{E.A} kept the period constant and set another scale (scale 2). The results of the new scale are shown in \fig\ref{fig:holens-case1}($E$). The higher hierarchy contains the lower hierarchy for the same region he selected. Furthermore, from the statistic view in \fig\ref{fig:holens-case1}($H$), he found that the proportion of the POI category has changed due to the change of movement modeling range after changing the scale. As shown in \fig\ref{fig:holens-case1}($D$) and \fig\ref{fig:holens-case1}($G$), the new movement aggregation range (higher hierarchy) inherits and expands the original movement aggregation range (lower hierarchy) by changing other scales under the same period.

\textbf{E.A} commented that the aggregation method could inspire him to make decisions on the planning of urban emergency response facilities. This method considers the contextual information to aggregate movements with more similar functions. He said that with this aggregation method, he can have an overview of what kind of activity is in the selected region. \textbf{E.A} gave an example: the function in the daytime is a transportation hub and business area, whereas, at night, the region tends to have more role in nightlife spots. Therefore, they can pay more attention to fire control and public security at night in such regions, according to the gained information, so as to reduce the risk of fire and public safety that recreational activities may bring about. Furthermore, the hierarchical multi-scale region segmentation method can provide intuitive representations to fit their analysis at different demands.

\subsection{Study 2: Interactive Exploration and Comparison }\label{sec: holens-usage scenario}

This case demonstrates the effectiveness of the higher-order pattern analysis (requirement \textbf{R.3}, \textbf{R.4}, and \textbf{R.5}), which aims to show how analysts explore and analyze the higher-order patterns (global/local) to make decisions. 

\vspace{1.5mm}
\noindent\textbf{Global Higher-order Pattern Exploration.}~\textbf{E.A} first selected holidays as the exploration day type, and it shows that the global higher-order patterns are all concentrated in the afternoon and evening in the temporal dimension (\fig\ref{fig:holens-teaser}($A_4$)). On the contrary, when \textbf{E.A} selected the weekdays, the higher-order patterns are distributed in the morning and the afternoon (\fig\ref{fig:holens-teaser}($A_2$)). \textbf{E.A} commented that human activity in urban areas usually starts early in the working days, whereas people usually choose to relax during holidays. For detailed exploration, when \textbf{E.A} clicked the patterns in the temporal distribution matrix one by one, he found that most of the patterns are associated with \textbf{R9} and \textbf{R13} (\fig\ref{fig:holens-teaser}($B_3$)), where these two regions are the downtown and midtown of Manhattan. 
In the state transition view (\fig\ref{fig: holens-case study2 global higher-order pattern}($B_1$)), the statistical result shows that the dominant activities for humans in \textbf{R9} are shopping and having meals (the proportion of arc). Moreover, \textbf{R13} is considerably different from \textbf{R9} which \textit{Outdoors \& Recreation} in \textbf{R13} accounts for a significant proportion. \textbf{E.A} mentioned that this is because \textbf{R13} includes Central Park. After he explored all the higher-order patterns in holidays and weekdays, it shows that although most global higher-order patterns are concentrated in Manhattan, the temporal dimensions of these patterns are different. In addition, the details of the function in these regions vary in different day types. \textbf{E.A} said this might increase their confidence to prioritize the planning of Manhattan to make a more effective urban area. Furthermore, \textbf{E.A} also summarized two different modes: the annular higher-order pattern (the relationship between two adjacent regions, \eg, \fig\ref{fig: holens-case study2 global higher-order pattern}($B_1$, $B_2$, and $B_3$ )) and the linear higher-order pattern (\eg, \fig\ref{fig: holens-case study2 global higher-order pattern}($B_4$). These modes vary because, on the city scale, many people’s activities can be completed inside the segmented region. Therefore, the global higher-order pattern mode is relatively simple.

\begin{figure}[htbp]
\centering
  \includegraphics[width=\linewidth]{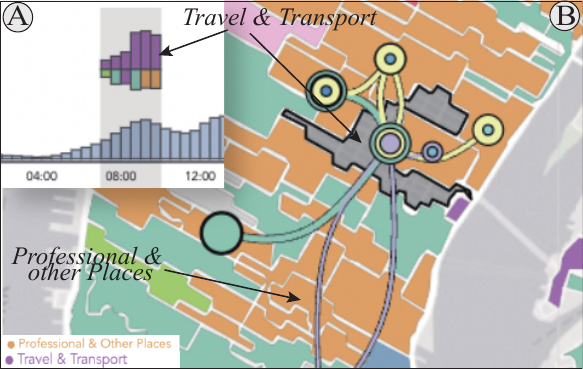}
    \vspace{-1.5em}
    \caption{The controlled timeline (A) and the selected region (B) perform the function of \textit{Travel \& Transport} when the period to 07:00-10:00.}
    \label{fig: teaser-case-2}
  \vspace{-0.5em}
\end{figure}
\noindent\textbf{Local Higher-order Pattern Exploration.}~
According to the global higher-order patterns, \textbf{E.A} selected the research area in Manhattan. He set the scale 3, as a default setting to explore the local higher-order pattern. He first slid the sliding window to 07:00-24:00 (\fig\ref{fig:holens-teaser}($A_3$)) to explore how the higher-order patterns happened in one working day. He arbitrarily selected one region (\fig\ref{fig:holens-teaser}($B_1$)) in midtown in Manhattan. From the bi-directional bar chart (\fig\ref{fig:holens-teaser}($A_3$)), \textbf{E.A} found that the majority of people check in for meals in this area, whereas the people who check in at the station dominate during the morning and evening commutes (\ie, commuting by public transportation in the morning and night). According to the H-Flow (\fig\ref{fig:holens-teaser}($B_1$)), \textbf{E.A} found that all corresponding regions are not connected with the selected regions. \textbf{E.A} said this is mainly because of the developed urban transportation, which people can use to commute to their destination far away. He then chose two patterns that share the same starting point (\fig\ref{fig:holens-teaser}($B_1$)). The two patterns both start at \textbf{R30884} and pass through \textbf{R30881}, but at the end, two different destinations appear. The edge (\fig\ref{fig:holens-teaser}($D_2$)) shows that the flow between the two regions is large in the morning and evening, respectively. \textbf{E.A} said that given the close distance between the endpoints, in combination with the flow between the two locations, it is an $A-B-A$ pattern. The reason the endpoints are different is that people rarely check in at home. The users usually check in outside areas such as dining, traveling, and shopping. Therefore, he was inspired by the urban planning task, which can further explore the two regions to optimize the commuting time in terms of public transportation scheduling between the two regions, such as increasing public transportation shifts between the two regions in the morning and evening rush hours. \textbf{E.A} then changed the period to 07:00-10:00 (\fig\ref{fig: teaser-case-2}($A$)).
As expected, the segmentation changes and the selected region performs the function of \textit{Travel \& Transport} (\fig\ref{fig: teaser-case-2}($B$)). Most of the surroundings (\fig\ref{fig: teaser-case-2}($B$))exhibit the function of \textit{Professional \& other Places}, which means a business function. In addition, the regions associated with this area are close. Combined with the in-flow from the timeline (\fig\ref{fig: teaser-case-2}($A$)), the transportation POI dominates in this period. \textbf{E.A} believes that this area reflects more traffic attributes in this period, that is, people come here through public transportation on weekdays and then work around this region.

\textbf{E.A} mentioned that the \holens~inspired his work on traffic management and urban planning. From the statistical view, he concluded that the selected region in the case study contains the main functions of \textit{Travel \& Transport}, \textit{Food}, and \textit{Professional \& Other Places}, and the dominant function changes the time, \ie, it functions as a transportation hub in the morning and evening, mainly serving as an office and catering during the other time. In addition, through the exploration of higher-order patterns, \textbf{E.A} found some related areas, such as \textbf{R30884} and \textbf{R30881}, where people are more willing to go after arriving at this place. As \textbf{E.A} is mainly engaged in urban safety and urban traffic planning, it can plan public transport routes in relevant areas in the morning and evening, such as increasing transport capacity and optimizing scheduling, and reasonably plan some urban emergency safety facilities, such as fire hoses and shelters, in case of disasters.

\subsection{Expert Interview}\label{sec: holens-expert-interview}
We interviewed three independent domain experts to evaluate and provide suggestions for this research. The first expert (\textbf{E.A}) is a research assistant professor at University S, whose research lies on urban computing. The second expert (\textbf{E.B}) is an assistant professor at University T, specializing in urban visualization and intelligent traffic. The third expert (\textbf{E.C}) is an analyst at a consulting firm with a degree in urban planning. \textbf{E.A} and \textbf{E.B} have Ph.D. degrees, and \textbf{E.C} has an MSc degree and over four years of experience in urban planning. Moreover, \textbf{E.B} and \textbf{E.C} have living experience in New York, which is the city focused on in the case study. First we briefly introduce our research and show them a demo video of the \holens. We then collect the feedback and summarize it from the aspect of feasibility, usability, and effectiveness. All the experts provided valuable feedback and suggestions according to their backgrounds. The summary of the interview is detailed below. 

\vspace{1.5mm}
\noindent\textbf{Feasibility \& Usability.}~The three experts agreed that the aggregation method in \holens~is a good attempt. The visual analytics on higher-order patterns based on the aggregation method enlightens urban planning. This study is beneficial for their research and work. \textbf{E.A} commented that ``the \holens~considers both the spatial proximity and the spatial contextual information when aggregating the data. The aggregation method is self-organized and adaptive, which fits the real demand when analyzing the higher-order pattern at different scales in the urban area". \textbf{E.B} mentioned that ``the research considers the temporal feature and segmented functional regions by considering dynamic access frequency rather than only static POI distribution, it is of great research value." They all agreed that our visual analytics on analyzing the higher-order patterns has significant reference value. \textbf{E.B} said that ``the visual effect of region segmentation on the map view is intuitive, the design of the H-Flow and the higher-order state sequence chart is useful on representing the higher-order patterns and can provide the function on visual comparison. I can follow the scenario easily, and the visual analytical process of the proposed method is feasible, which can truly help in urban decision-making."

\vspace{1.5mm}
\noindent\textbf{Effectiveness.}~All three experts agreed that the visual analytics approach is effective for their work. \textbf{E.B} said that ``the \holens~is enlightening to my current research on infectious disease prevention and control. Your method can segment the region according to the real access situation, and the range of the region segmentation is dynamic considering the temporal variability. Moreover, we can learn from the extracted higher-order patterns that those people who appear in a specific place at a specific time, where do most of them come from and are more likely to go". \textbf{E.C} shared his project experience for optimizing the taxi route planning strategy. The project is to provide strategies for taxi companies to reduce the no-load rate of taxis. They applied the algorithm based on the Monte-Carlo search tree. However, the method requires a rich mathematical foundation to understand the model. In addition, their model hardly provides strategy under different scales and cannot provide more detailed information on the customers, \eg, whether the customer takes a taxi after drinking. Moreover, \textbf{E.C} said that \holens~could improve the interpretation according to the interaction with the UI and provide an intuitive visualization of what the people in the current location tend to do and where they tend to go next. 

\vspace{1.5mm}
\noindent\textbf{Suggestions.}~The experts provided fruitful suggestions on the design and the potential usage directions. \textbf{E.A} and \textbf{E.C} commented that the design of the statistic view could be improved. The initial design of the statistic view just represents the statistical information of the access frequency; nevertheless, the experts reminded us that visualizing the POI statistic of the selected region is also essential, and we can make a detailed comparison. Furthermore, \textbf{E.A} said that we need to better distinguish between the origin and destination of the higher-order patterns because this is crucial in an annular higher-order pattern mode. Therefore, we used tornado diagrams in the statistic view to represent the statistic information of POI categories and POI access frequency. In addition, a small circle in the last node is nested to represent the entropy rate of the higher-order flow and bold the origin node better to identify the origin and destination of the higher-order pattern. \textbf{E.B} combines his current research and gives a possible application scenario. He suggests that we can integrate other spatial-temporal urban data, \eg, telco data, which can be used for epidemiological investigation. The self-organized aggregation method can segment the risk area according to the spatial context and the people's accessibility, not just the spatial proximity. In addition, the higher-order pattern exploration pipeline can help analysts predict the potential activity scope of the infectious source and infer the region where the infectious source may have passed in the past. Moreover, all the experts said that the computational efficiency of the modeling algorithm could be improved.


\section{Discussion}\label{sec: holens-discussion}
This section discusses the parameter selection (\sect\ref{sec: holens-parameter selection}) and the determination of order number~(\sect\ref{sec: holens-order number}). Then we point out the limitations from the data perspective~(\sect\ref{sec: holens-data persective}) and some inspirations of the multi-variant visualization~(\sect\ref{sec: holens-Multi-variant Visualization}).

\subsection{Parameter Selection}\label{sec: holens-parameter selection}
In this research, two parameters determine the performance and the visual effect, that is, the aggregating threshold $\alpha$ and the aggregation range $\beta$. $\alpha$ controls the threshold of the entropy when aggregating the regions, which is the scale of the aggregation. Intuitively, the smaller $\alpha$, the easier the aggregation. $\beta$ is to control the aggregation process under a reasonable spatial range, making the visual effect of the aggregation neither too big nor too small in size. The parameter selection may affect users’ decision-making. 

\begin{figure}[htbp]
\centering
  \includegraphics[width=0.95\linewidth]{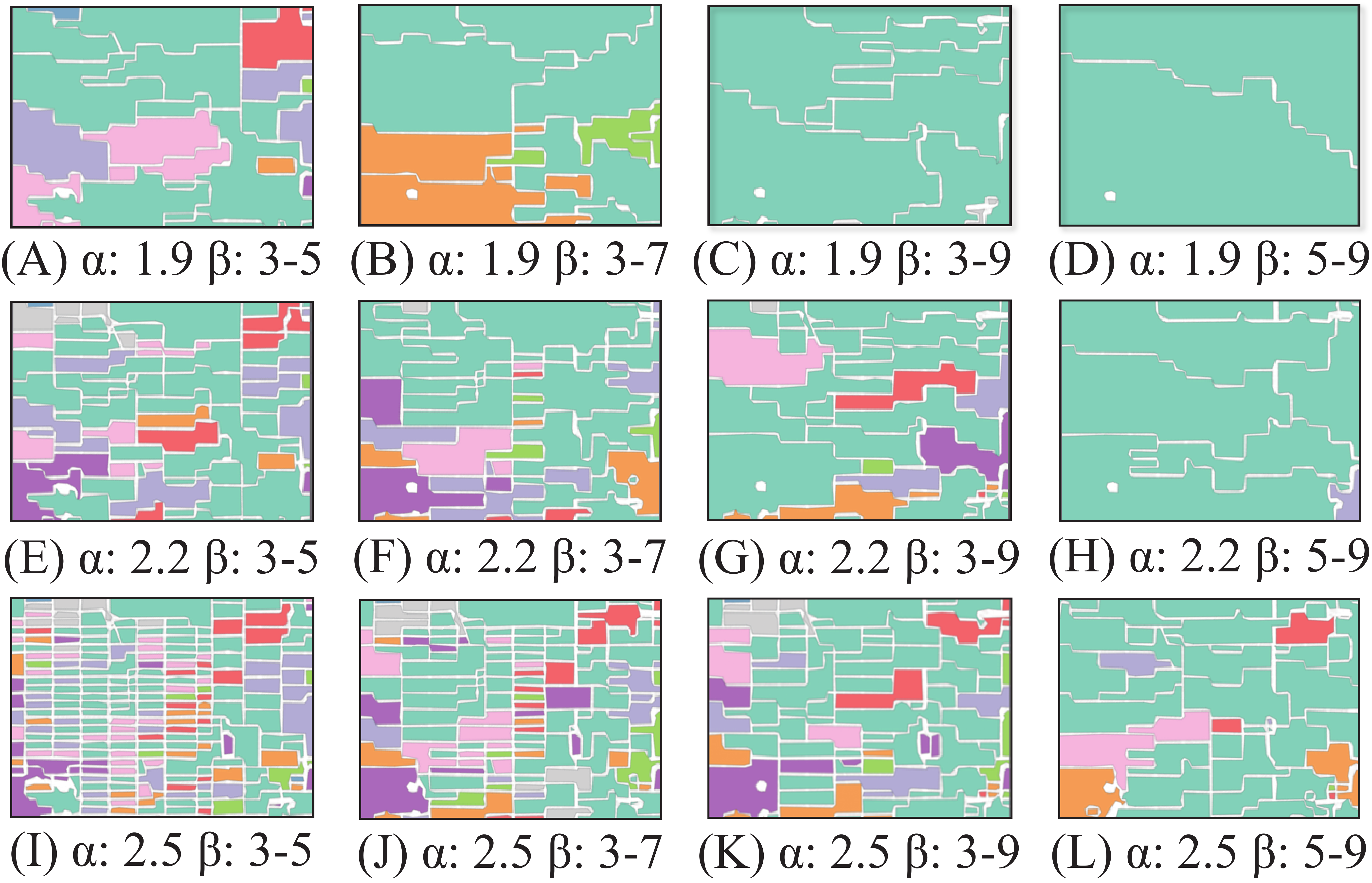}
  \vspace{-0.5em}
  
  \caption{Parameter selection: (A) to (D) are the aggregating range $\beta$ with different range with the aggregating threshold $\alpha$ = 1.9;  (E) to (H) are the aggregating range $\beta$ with different range with the aggregating threshold $\alpha$ = 2.2; (I) to (L) are the aggregating range $\beta$ with different range with the aggregating threshold $\alpha$ = 2.5.}
  \label{fig:holens-discussion-range-scale}
   \vspace{-1em}
\end{figure}

\fig\ref{fig:holens-discussion-range-scale} is the different combinations of the $\alpha$ and $\beta$. We select one region in \textbf{R13} as the AOI, and the exploration period is 18:00-20:00. We select 3-5, 3-7, 3-9, and 5-9 as the candidates of $\beta$. As shown in \fig\ref{fig:holens-discussion-range-scale}, the different combinations have different visual effects. In \fig\ref{fig:holens-discussion-range-scale}($A$, $E$, $I$), the aggregation effect is not evident when the scale is small (\ie, $\alpha$ = 2.5). A similar problem is also in \fig\ref{fig:holens-discussion-range-scale}($J$), where $\beta$ is 3-7. Compared with \fig\ref{fig:holens-discussion-range-scale}($C$,$G$,$K$) ($\beta$ = 3-9 ), it can be summarized that if the upper limit of $\beta$ is too small, the aggregation will not be evident when $\alpha$ is relatively large. However, when the upper limit is reasonable, the lower limit leads to a large aggregation when $\alpha$ is small (\eg, \fig\ref{fig:holens-discussion-range-scale}($D$)) because when $\alpha$ is large, the aggregation result is relatively large, and the method holds that the larger hierarchy will aggregate on the basis of the previous hierarchy. Moreover, when the difference between the upper and lower limit is too large, the aggregation of the same hierarchy may significantly differ.

The spatial range ensures that the aggregated region will not become extreme due to the setting of $\alpha$, that is, the area of the aggregated regions varies. This variation may mislead the decision-making when the size of the segmented region varies greatly and exists on a map simultaneously. This study eventually selects 3-9 as the $\beta$, by considering the visual effects on the selected regions. In the case studies, we focus on exploring midtown and downtown Manhattan, where data is most abundant. In addition, the blocks in this region are divided into similar areas. However, in other city regions, \eg, in Brooklyn, the blocks are not divided regularly; thus, the aggregation range is difficult to control flexibly. Therefore, finding a reasonable parameter $\beta$ to adapt to the regular blocks in Manhattan and the irregular blocks in other areas simultaneously is difficult. Moreover, when the region is aggregated at different scales, a larger scale is aggregated on the basis of a smaller scale. Therefore, ensuring that all scales remain stable in terms of visual effect by setting one aggregating range $\beta$ is also challenging. Thus, $\beta$ = 3-9 is not the best one, which just seems reasonable in our case studies. An algorithm that can adaptively control the aggregation range according to the size of the aggregation region is necessary for the future, but it is not involved in this work.

\subsection{Order Number Determination}\label{sec: holens-order number}

In this research, the ``higher-order" is a multi-step propagation. However, the number of the order is a complicated problem that needs long pathways to distinguish the real effects due to the lack of data~\cite{chierichetti2012web}. Previous studies have had many different viewpoints \eg, Rosvall et al.~\cite{rosvall2014memory} mentioned that second-order is statistically significant on ranking and spreading dynamics, whereas Tao et al.~\cite{tao2017honvis} claimed that it could depend on up to five. Our research applies the KLD to measure the probability distribution changes and decide whether the extraction of the higher-order patterns stops. However, these criteria depend on our threshold setting. 
\begin{figure}[htbp]
\centering
  \includegraphics[width=\linewidth]{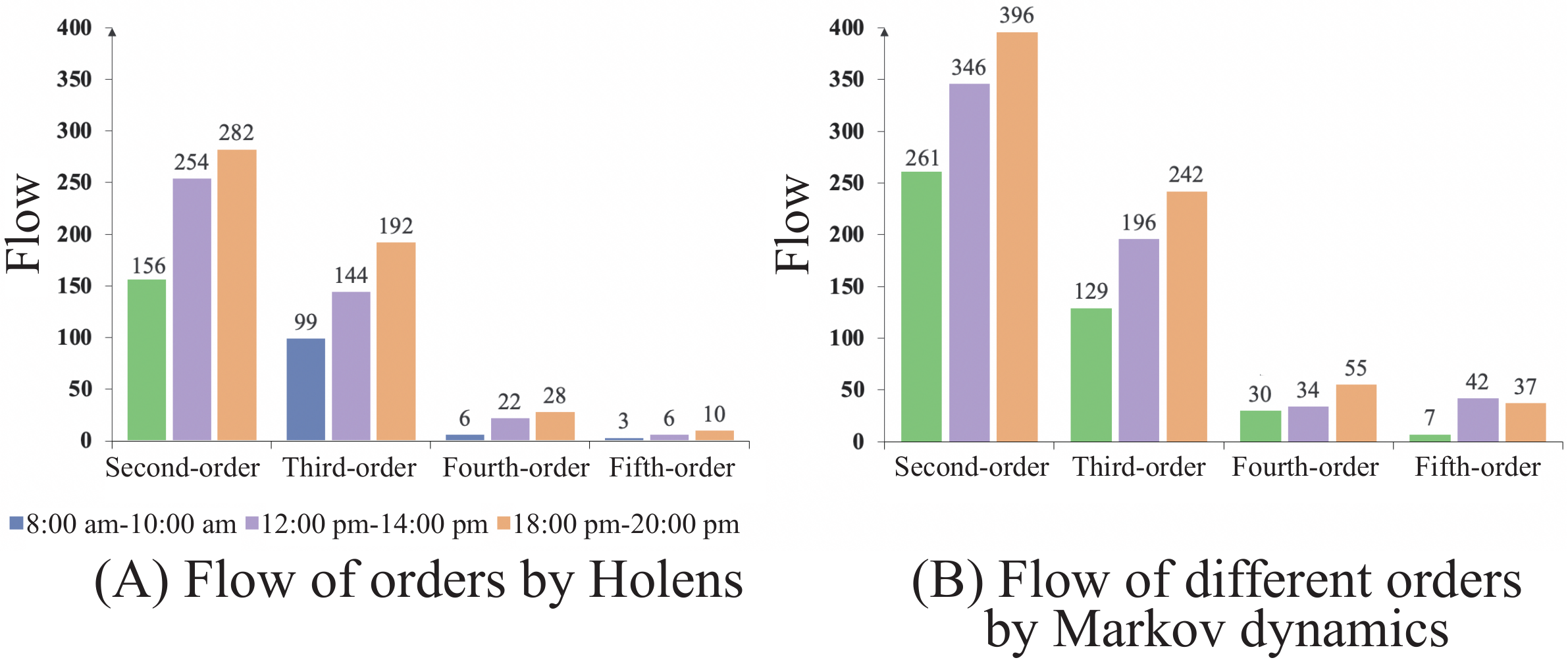}
    \vspace{-1.5em}
    \caption{The results of average flow statistics of higher-order patterns of different numbers of orders by (A) our proposed method and by (B) Markov dynamics.}
    \label{fig: holens-disscussion-order-num}
  \vspace{-0.5em}
\end{figure}
We select these periods (7:00-10:00, 12:00.-14:00, and 18:00-20:00) and several arbitrary regions in \textbf{R13} to explore the number of orders. We count the number of flows in each pattern to observe the statistical effect on different orders. \fig\ref{fig: holens-disscussion-order-num} shows the results of average flow statistics of higher-order patterns of different numbers of orders by both the method and the Markov dynamics~\cite{tao2017honvis}. We found that when the number of orders increases to four in both methods, a significant decline occurs in the flow in the higher-order pattern. That is, as the number of orders increases, each higher-order pattern becomes unique. Therefore, the flow with the same trajectory in a certain period decreases. To meet the real situation, \holens~also considers the flow volume. If the higher-order pattern contains few trajectories, it still does not consider a higher-order pattern. And according to \fig\ref{fig: holens-disscussion-order-num}, the order numbers are limited to 3 at most, which means we only focus on the second-and third-order in \holens. Nevertheless, during this research, an automatic algorithm for deciding the number of orders is still an urgent demand in future exploration.

\subsection{Data Perspective}\label{sec: holens-data persective}
This work uses the check-in data from~\cite{yang2019revisiting,yang2020lbsn2vec++}. This dataset labels what the people are doing at that time in the checked timestamp. It is a digital footprint for the people. 
In contrast to the taxi trajectory data, which the sensor may sample at regular temporal intervals, the human movement check-in data are sampled by humans; thus, the temporal interval for the data is random, but each sampling point has real meaning. In this research, the analysts did not conduct a one-day or real-time analysis of the traffic management and urban planning tasks. Only two day types need to be considered: weekdays and weekends. Thus, when we overlay these data into two different day types, the dataset is relatively dense. However, these data form sequences of points but not a series of continuous stay periods. Therefore, we deduce the stay for each person at one location by the shortest path (network distance) between two consecutive check-in points. We ignore their commuting time if the two consecutive points are in the same building. Humans have various ways to commute between two check-in points, \ie, by vehicle or walking, and some may use public transportation. Although one cannot acquire the accurate commuting time when one person moves inside one building, the deduced stay period can be just a reference, which is reasonable to a certain extent and cannot be guaranteed to be exactly correct.

In addition, as shown in \fig\ref{fig:holens-teaser}($B$), we found some blank areas without any POI label in the map view because there is no check-in data from the dataset in these regions, which is the limitation of the data itself and will be further improved with the improvement of data quality. Furthermore, the label of the dataset may affect the analysis of the higher-order pattern in the urban area. 
The users tend to check in at public places, \eg, restaurants, metro stations, gyms, shops, etc. However, seldom of them check in at home, or at best, they check in at the station near their home, resulting in difficulty in analyzing the higher-order pattern behavior on human's commuting, \eg, the pattern home$\rightarrow$ working place$\rightarrow$ home. 
We believe that these limitations will be solved with the improvement of data richness and data quality.

\subsection{Multi-variant Visualization}\label{sec: holens-Multi-variant Visualization}
\holens~uses a color scheme to represent the POI categories.
The dataset contains nine different POIs; thus, it can select nine from the color library to represent different POI categories. 
However, if there are too many types of POI, the current representation will become unreasonable. Given that too many colors are displayed on the map, the interface is chaotic. In addition, increasing colors leads to visual proximity between different colors, thus affecting analysts' decision-making. For such multi-variant visualization problems, several researchers (\eg,~\cite{cao2011dicon,yang2007value,post1995iconic,sun2020dfseer}) have focused on this. In DFSeer~\cite{sun2020dfseer}, the authors used colors to represent different categories of ML models. However, this work chooses 5 of the 22 models to visualize. Therefore, it only needs to deal with the selected five models and gives each selected model a color, which does not need to be consistent. This is not suitable for my demand. My solution seems to be relatively suitable for the current dataset and our current demands. Moreover, the multi-variant visualization problem on a map remains a challenge to the VIS community.


\section{Conclusion}\label{sec: conclusion}

In this study, we proposed \holens, a comprehensive visual analytics approach for analyzing higher-order patterns in the urban area, including a novel aggregation method and a web-based visual interface. The aggregation method in \holens~outperforms conventional urban movement methods in terms of rationality and visual effect of the conventional movement aggregation methods in satisfying the requirement of adaptively aggregating the urban movement data by considering the spatial proximity and the urban contextual information. Moreover, \holens~ supports hierarchical organizing, 
indicating that the analysts can analyze the higher-order pattern at different scales. The visual design of \holens~satisfies the requirement of visualizing the movement aggregation, higher-order pattern extraction, visualization, and comparison. \holens~supports exploring the higher-order pattern in terms of temporal dimension by improving the algorithm from the conventional approach. 
\holens~consists of several novel designed visualization techniques \ie, H-Flow and higher-order state sequence chart. In this study, we consider the different combinations of the parameters and the alternative designs, and we choose an optimal selection with good performance on intuitive visualization. Finally, two real-world case studies and the feedback from an interview with domain experts demonstrate the feasibility, usability, and effectiveness of \holens.

\subsection*{Acknowledgements}
We thank all the domain experts interviewed in this study. We also thank the reviewers for their comments and suggestions. This work was supported in part by the Shenzhen Science and Technology Program (Grant No. ZDSYS20210623092007023) in part by National Natural Science Foundation of China (No. 62172398) and Guangdong Basic and Applied Basic Research Foundation (2021A1515011700).
\subsection*{Declarations}
There is no conflict of interest/competing Interest in this research.

\bibliographystyle{CVMbib}
\bibliography{refs}

\subsection*{Author biography}

\begin{biography}[{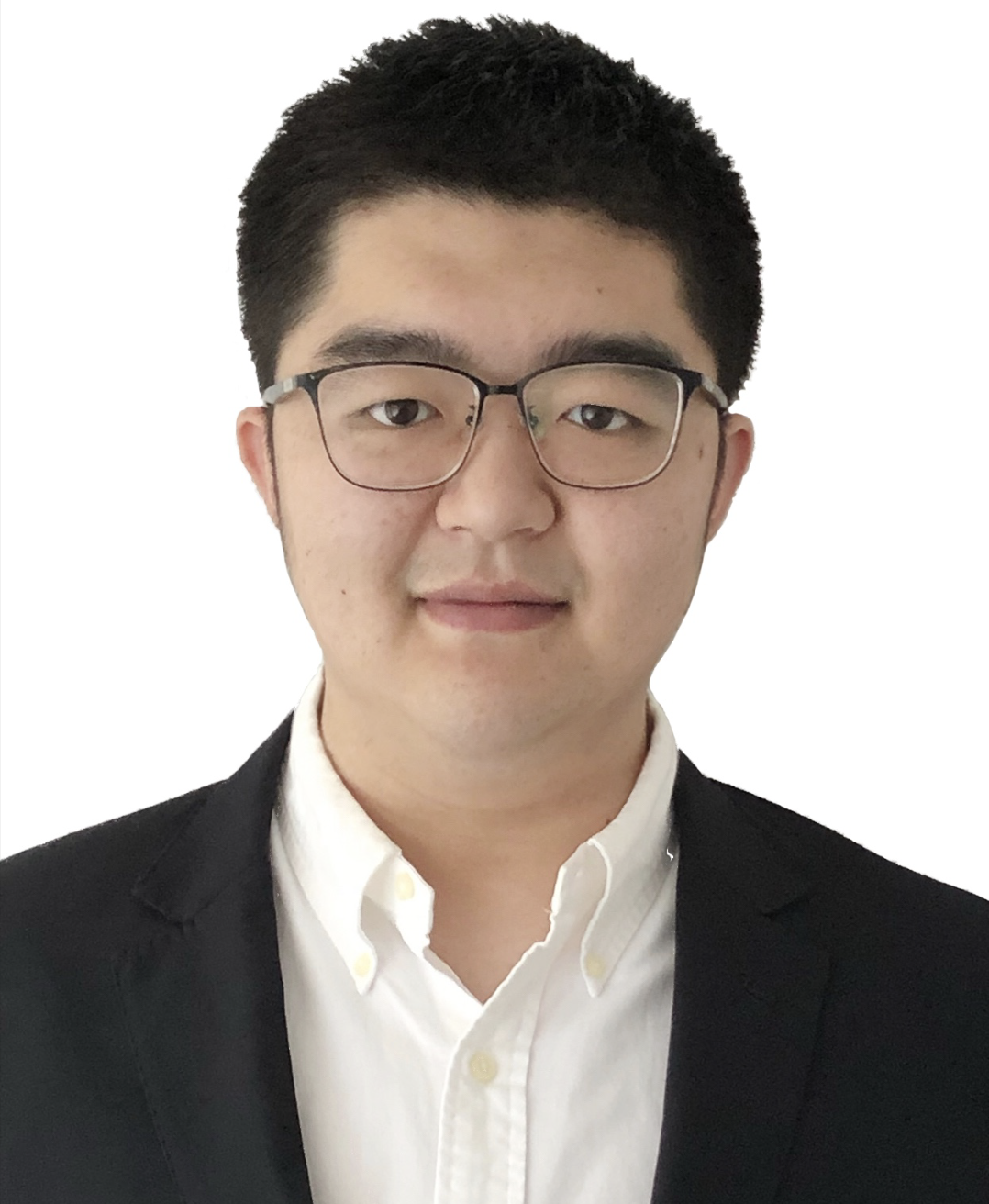}]{Zezheng Feng} received a B.E. degree from Northeastern University (NEU), China, in 2017, and an M.S. degree (with distinction) from Loughborough University, UK, in 2018. He is currently a Ph.D. candidate in the Department of Computer Science and Engineering (CSE) at the Hong Kong University of Science and Technology (HKUST) and sponsored by a joint Ph.D. program between HKUST and Southern University of Science and Technology (SUSTech). His recent research interests include visualization and visual analytics, explainable artificial intelligence (XAI), and urban computing. For more information, please visit \url{https://jerrodfeng.github.io/}
\end{biography}
\vspace*{2em}
\begin{biography}[{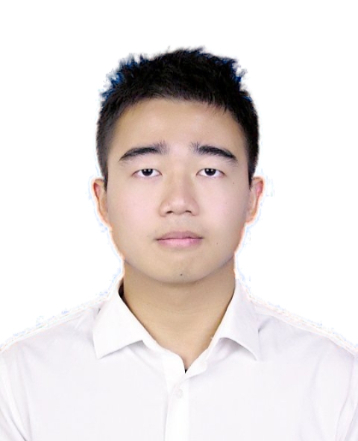}]{Fang Zhu}
received a B.E. degree in computer science and technology from the Southern University of Science and Technology, China, in 2022. He is working toward an M.S. degree from the Southern University of Science and Technology. His research interests include visual analytics and explainable AI.
\end{biography}
\vspace*{2em}
\begin{biography}[{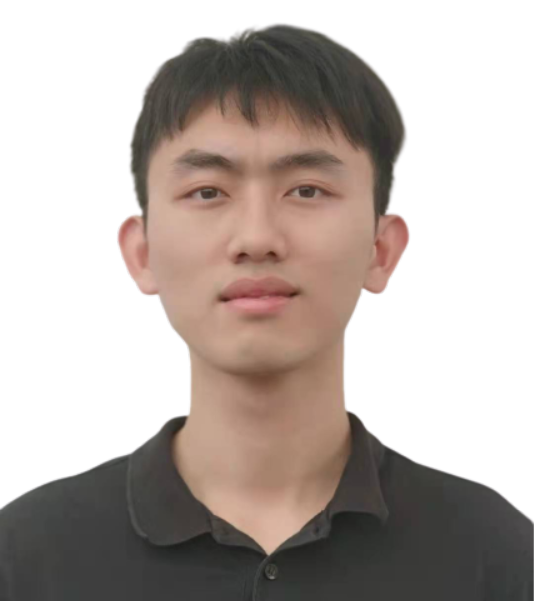}]{Hongjun Wang} 
is working toward an M.S. degree in computer science and technology from the Southern University of Science and Technology, China. He received a B.E. degree from the Nanjing University of Posts and Telecommunications, China, in 2019. His research interests are broadly in machine learning, urban computing, explainable AI, data mining, and data visualization.
\end{biography}
\vspace*{2em}
\begin{biography}[{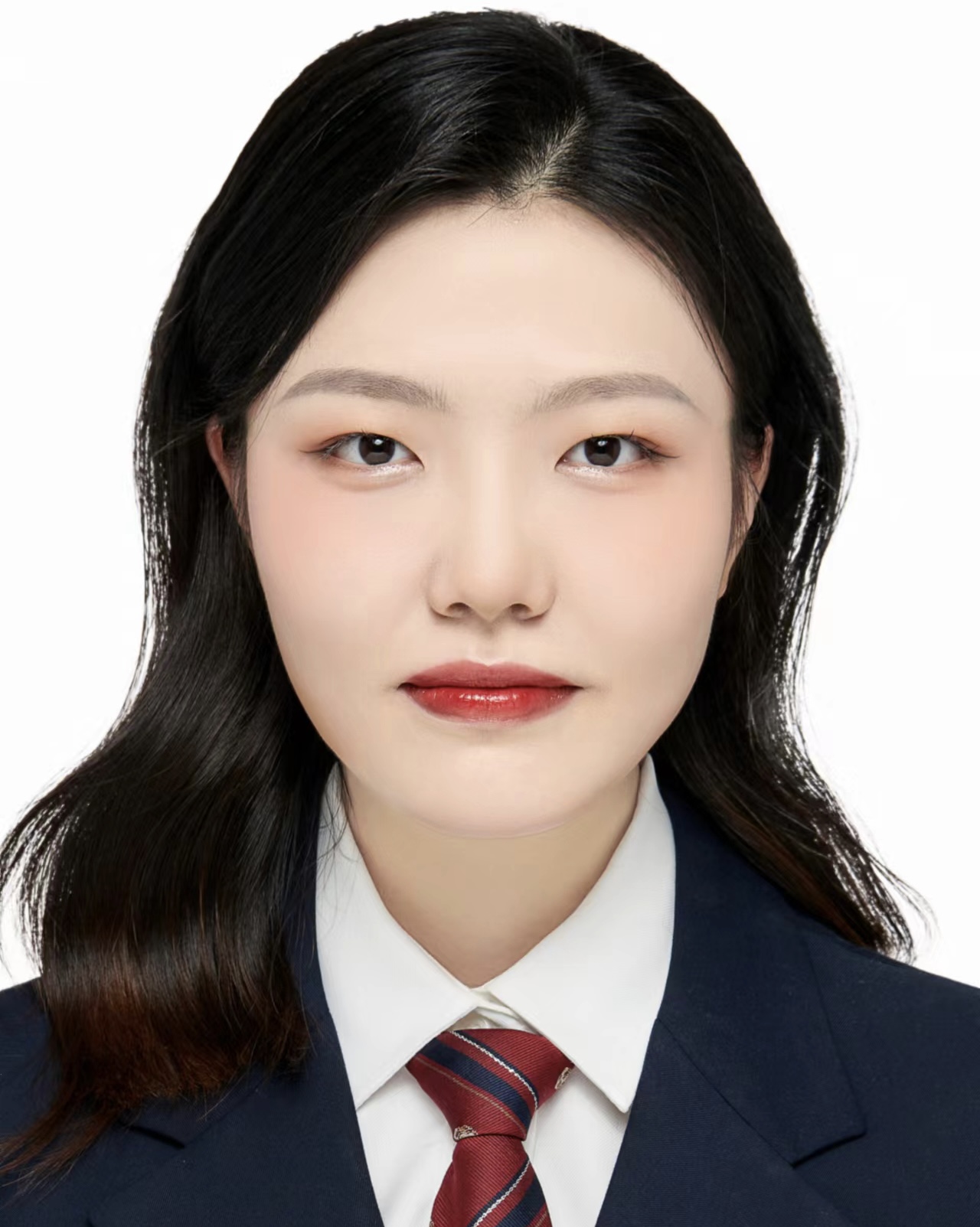}]{Jianing Hao} 
received a B.E. degree in computer science and technology from Shandong University, China, in 2022. She is currently a Ph.D. student at The Hong Kong University of Science and Technology (Guangzhou). Her research interests include Human-AI collaboration and visual analytics.
\end{biography}
\vspace*{2em}

\begin{biography}[{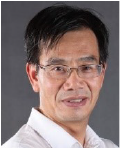}]{Shuang-Hua Yang} 
received the B.S. degree in instrument and automation and the M.S. degree in process control from the China University of Petroleum (Huadong), Beijing, China, in 1983 and 1986, respectively, and the Ph.D. degree in intelligent systems from Zhejiang University, Hangzhou, China, in 1991. He is currently the director of the Shenzhen Key Laboratory of Safety and Security for Next Generation of Industrial Internet at the Southern University of Science and Technology, China, and also the Head of Department of Computer Science at the University of Reading, UK. His research interests include cyber-physical systems, the Internet of Things, wireless network-based monitoring and control, and safety-critical systems. He is a fellow of IET and InstMC, U.K. He is also an Associate Editor of IET Cyber-Physical Systems: Theory and Applications.
\end{biography}
\vspace*{2em}
\begin{biography}[{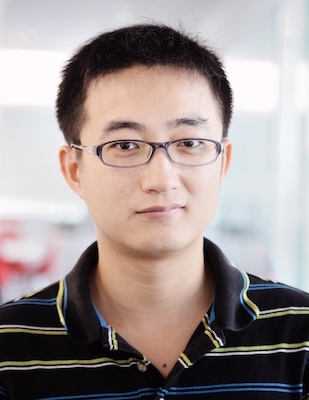}]{Wei Zeng} 
is an assistant professor in the Hong Kong University of Science and Technology (Guangzhou).
He received the PhD degree in computer science from Nanyang Technological University, and worked as a senior research at Future Cities Laboratory, ETH Zurich, and an associate researcher at Shenzhen Institute of Advanced Technology, Chinese Academy of Sciences. His recent research interests include visualization and visual analytics, computer graphics, AR/VR, and HCI. For more information, please visit \url{https://zeng-wei.com/}
\end{biography}
\vspace*{2em}

\begin{biography}[{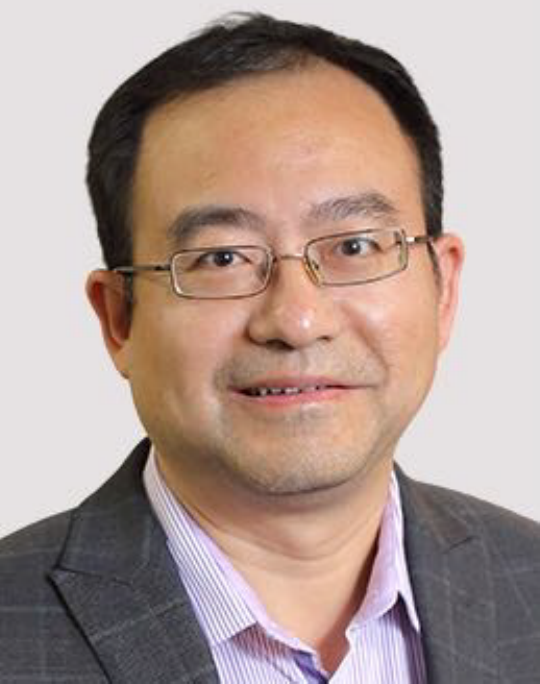}]{Huamin Qu} 
is a professor in the Department of Computer Science and Engineering (CSE) at the Hong Kong University of Science and Technology (HKUST) and also the director of the interdisciplinary program office (IPO) of HKUST. He obtained a BS in Mathematics from Xi'an Jiaotong University, China, an MS and a PhD in Computer Science from the Stony Brook University. His main research interests are in visualization and human-computer interaction, with focuses on urban informatics, social network analysis, E-learning, text visualization, and explainable artificial intelligence (XAI). For more information, please visit \url{http://huamin.org/}.
\end{biography}
\vspace*{2em}

\end{document}